\documentclass[preprint,pra,onecolumn,12pt]{revtex4}%
\usepackage{amsfonts}
\usepackage{amsmath}
\usepackage{amssymb}
\usepackage{graphicx}
\usepackage{dcolumn}%
\providecommand{\U}[1]{\protect\rule{.1in}{.1in}}

\begin{document}
\preprint{ }
\title{The double Caldeira-Leggett model: Derivation and solutions of the master
equations, reservoir-induced interactions and decoherence}
\author{A. Cacheffo$^{1}$, M. H. Y. Moussa$^{2}$, and M. A. de Ponte$^{1}$}
\affiliation{$^{1}$Departamento de F\'{\i}sica, Universidade Federal de S\~{a}o Carlos,
Caixa Postal 676, S\~{a}o Carlos, 13565-905, S\~{a}o Paulo,\textit{ }Brazil}
\affiliation{$^{2}$ Instituto de F\'{\i}sica de S\~{a}o Carlos, Universidade de S\~{a}o
Paulo, Caixa Postal 369, 13560-590 S\~{a}o Carlos, SP, Brazil }

\begin{abstract}
In this paper we analyze the double Caldeira-Leggett model: the path integral
approach to two interacting dissipative harmonic oscillators. Assuming a
general form of the interaction between the oscillators, we consider two
different situations: i) when each oscillator is coupled to its own reservoir,
and ii) when both oscillators are coupled to a common reservoir. After
deriving and solving the master equation for each case, we analyze the
decoherence process of particular entanglements in the positional space of
both oscillators. To analyze the decoherence mechanism we have derived a
general decay function for the off-diagonal peaks of the density matrix, which
applies both to a common and separate reservoirs. We have also identified the
expected interaction between the two dissipative oscillators induced by their
common reservoir. Such reservoir-induced interaction, which gives rise to
interesting collective damping effects, such as the emergence of relaxation-
and decoherence-free subspaces, is shown to be blurred by the high-temperature
regime considered in this study. However, we find that different interactions
between the dissipative oscillators, described by rotating or counter-rotating
terms, result in different decay rates for the interference terms of the
density matrix.

\end{abstract}

\maketitle

\section{Introduction}

At the beginning of the 1980s, the work of Zurek \cite{Zurek}, Caldeira and
Leggett (CL) \cite{CL}, and Zeh and Joos \cite{ZJ} played a decisive role in
the understanding of the still unsolved phenomenon of quantum measurement;
more specifically, the collapse of the wave function and the associated
decoherence of superposition states \cite{PRA}. Taking the reservoir into
account explicitly as a quantum ingredient, and analyzing its effect on the
evolution of an initial pure state into a statistical mixture, these papers
shed light on the shadowy interface between microscopic and macroscopic
domains. Although the wave function collapse remains an obscure process,
despite striking contributions also dating from the eighties \cite{GRWP}, much
is known today about the mechanisms leading to decoherence. In the last few
decades we have analyzed this phenomenon exhaustively, enabling the
proposition of a plethora of protocols to circumvent it, ranging from quantum
error correction codes QECC \cite{QECC} and engineered reservoirs \cite{ER} to
dynamical decoupling \cite{DD} and relaxation- and decoherence-free subspaces
(R-DFSs) \cite{DFS,Mickel-AP2}.

More recently, it was demonstrated that entanglement shows scaling behavior in
the vicinity of the transition point \cite{Osterloh}. This connection between
the theories of critical phenomena and quantum information, together with the
search for R-DFSs --- which encompasses dissipative coupled systems --- has
triggered the study of fundamental quantum processes in the domain of
many-body physics. Apart from the crucial role played by entanglements in the
understanding of quantum phase transitions \cite{Vidal}, the study of the
complex dynamics of coherence and decoherence of superposition states in
networks of dissipative quantum systems has also produced interesting results
for quantum information theory \cite{Mickel-AP2,Mickel-AP1,Mickel}. In
particular, in Ref. \cite{Mickel-AP2} a correlation function was introduced to
provide the analytical conditions for the existence of R-DFSs in a system of
interacting dissipative resonators.\ This correlation function measures the
reliability of a decoherence-free subspace. Apart from the correlation
function, Refs. \cite{Mickel-AP2,Mickel-AP1,Mickel} put forward, as a
conceptual novelty, the need to consider distinct reservoirs for distinct
quantum systems.

In this study, we consider a system of two interacting harmonic oscillators,
in two situations: $i)$ when each one is coupled to its own reservoir and
$ii)$ when both oscillators are coupled to a common reservoir. As argued in
Ref. \cite{Mickel-AP2}, the former case, where each system interacts with its
own reservoir, is the most usual situation. Considering, for example, a
network of coupled cavities, even when they have the same quality factor their
damping mechanisms are independent, except when they interact strongly
\cite{Mickel-AP2}. By strong interaction it is meant that the coupling
strength $\lambda$ between the $N$ cavities composing the network must satisfy
the relation $N\lambda\sim\omega$, $\omega$ being the natural frequency of
each cavity mode in the degenerate case, where $\omega_{1}=\cdots=\omega
_{N}=\omega$. In such a strong coupling limit, the results emerging in the
case of a common reservoir are completely similar to those for the case where
distinct reservoirs are assumed. Conversely, in the weak coupling limit, where
$N\lambda\ll\omega$, separate reservoirs must be assigned to each cavity
composing the network.

Even in particular cases where we could, in principle, assign a common
reservoir to different quantum systems, such as a sample of atoms inside the
same cavity --- the multimodal cavity playing the role of a common reservoir
--- such a reservoir turns out to act as distinct reservoirs when the
transition frequencies of the atoms are significantly far from each other. In
this situation, each atom interacts with the reservoir modes around its own
frequency transition, and the absence of (or small) intersection between the
reservoir modes addressed by distinct atoms makes the common reservoir act as
several distinct ones \cite{Mickel-AP2}. Evidently, when the atomic transition
frequencies are not sufficiently apart from each other, the overlap between
the reservoir modes addressed by each atom indicates that they start to
interact through their \textquotedblleft partially
common\ reservoir\textquotedblright. Only in the limiting case, where all the
atoms in the sample have the same transition frequency --- and a full overlap
between the reservoir modes is achieved --- is a \textquotedblleft completely
common\ reservoir\textquotedblright\ accessed by the whole atomic sample
\cite{Mickel-AP2}. At this limit, the interaction between the atoms through
their common reservoir is maximized, opening up the possibility of an
interesting feature arising from interacting dissipative systems: the R-DFSs.
Therefore, the assessment of such R-DFSs --- which may become indispensable
for the implementation of quantum information --- requires a completely
understanding of the dissipative mechanism for coupled quantum systems: either
through a common or distinct reservoirs. Apart from the emergence of R-DFSs,
the subject of collective damping effects has recently produced interesting
results, such as the nonadditivity of decoherence rates observed in a network
of dissipative oscillators \cite{Mickel-AP2,Mickel-AP1,Mickel}, as well as in
superconducting qubits \cite{Brito}. Returning to the atomic samples, we
stress that the collective damping effects coming from two-atom systems
\cite{Ficek}, can be directly identified with the nonadditivity of decoherence rates.

The problem of two coupled harmonic oscillators has already been discussed in
the literature from various perspectives. We first mention a proposal, based
on the possibility of performing a reversible coupling between high-Q
cavities, to achieve reversible decoherence of a mesoscopic superposition of
field states \cite{Raimond}. A theoretical approach for such an experimental
proposal is given in Ref. \cite{Nemes}, where a common reservoir is assumed
for both cavities. In Ref. \cite{Zoubi} a system of two coupled cavities has
been analyzed in which just one of the cavities interacts with a reservoir. A
master equation is derived for the case of strongly coupled cavities and it is
shown that the relaxation term is not simply the standard one, obtained by
neglecting the interaction between the cavities. Finally, in Ref.
\cite{Mickel-AP1}, each cavity is assumed to interact with its own reservoir
and a detailed investigation is carried out for both regimes of weakly and
strongly coupled oscillators. In order that the oscillators interact only
through their direct coupling and not indirectly through their couplings with
a common reservoir, it is advisable to assume two distinct reservoirs. It is
worth mentioning that a general treatment of a network of coupled dissipative
harmonic oscillators has recently been presented \cite{MickelRG} for any
topology --- i.e., irrespective of how the oscillators are coupled together,
the strength of their couplings, and their natural frequencies. As in Refs.
\cite{Mickel-AP2,Mickel-AP1,Mickel}, the authors start with a general, more
realistic, scenario where each oscillator is coupled to its own reservoir, and
proceed later to the particular case where all the network oscillators are
coupled to a common reservoir.

In the present paper, a general form for the interaction between two
dissipative oscillators is considered and both situations, of distinct
reservoirs and a common one, are analyzed. Moreover, instead of the master
equation approach, we follow the path integral approach adopted by CL in their
linear response model \cite{CL}, with which considerable progress has been
made on the subject of quantum dissipation in several areas of physics. In
fact, with their linear response model, CL accounted for the influence of
dissipation on quantum tunneling in macroscopic systems \cite{Tunneling}.
Quantum Brownian motion has also been approached through the
influence-functional method \cite{CL,BrownianGrabert,BrownianZ}. Moreover, the
linear response model in \cite{CL} has been applied to many topics in solid
state physics, for example the dynamics of polarons \cite{Neto} and a particle
coupled to a Luttinger liquid \cite{Luttinger}.

Long before the CL model, functional integral calculations were used by
Feynman to analyze the problem of polarons in a polar crystal \cite{Feynman}
and, more recently, they have been applied to the problem of bipolarons
\cite{Wilson}. As a direct extension of the Feynman polaron model, in the
path-integral approach to the bipolaron each electron is harmonically coupled
to a fictitious heavy particle which replaces the virtual phonon cloud. Each
electron also interacts with the fictitious particle of the other electron,
apart from the Coulomb repulsion between the two. Therefore, the double CL
model presented here bears some resemblance to the bipolaron problem, with our
oscillators (reservoirs) replacing the electrons (fictitious particles). By
replacing the interaction between the two dissipative oscillators by the
Coulomb repulsion between the electrons, we end up with a dissipative
bipolaron-type model. In fact, as we demonstrate here, the oscillators
interacts indirectly through their common reservoir or even their separate ones.

Before closing this Introduction, we must mention the recent result in Ref.
\cite{Amir}, where the authors point out the effective coupling that is
induced between two Brownian noninteracting particles by a common reservoir.
Such an effective coupling depends on the choice made for the spectral
function of the reservoir. In the present study, working with an ohmic
reservoir at the high-temperature limit, we find that this induced effective
coupling occurs in the case of a common reservoir while, as expected, it is
absent in the case of distinct reservoirs. We also note that such an effective
coupling induced by a common reservoir is also pointed out in Refs.
\cite{Mickel-AP2,Mickel-AP1,Mickel,MickelRG}.

Summarizing, in the present paper we employ the path integral approach to
treat a network of two interacting dissipative harmonic oscillators. Assuming
a general form of the interaction between the oscillators, we consider two
different situations: i) when each oscillator is coupled to its own reservoir,
and ii) when both oscillators are coupled to a common reservoir. We derive and
solve the master equation for each case and, in the latter, we identify the
reservoir-induced coupling between the oscillators, which arises even when the
original interaction between them is schwitched off. We verify that such a
reservoir-induced coupling encompasses both dissipative and diffusive terms
which couple together the variables of both oscillators. These terms thus
account for the energy loss of the oscillators through each other, apart from
a joint diffusive process. From the solutions we have found for the master
equations we also compute a general expression for the decay rate of the
off-diagonal peaks of the density matrix of initial superposition states,
which also applies to both cases of a common and distinct reservoirs. Finally,
considering different interactions between the oscillators, we analyze the
decoherence process of particular entanglements in their positional spaces.

\section{The Double Caldeira-Leggett Model}

The CL problem consists in applying the influence-functional method of Feynman
and Vernon to study the dissipation-fluctuation dynamics of a quantum system
($S$) interacting with a reservoir\ ($R$) modelled by a collection of
one-dimensional harmonic oscillators. The CL Hamiltonian is given by
$H_{CL}=H_{S}+H_{R}+H_{S/R}$, where the Hamiltonian of the system, represented
by a particle of mass $m$ and coordinates $q$ and $p$, is given by
\begin{equation}
H_{S}=\frac{p^{2}}{2m}+V(q)\text{;} \label{1}%
\end{equation}
the Hamiltonian for the reservoir, consisting of a collection of harmonic
oscillators $\left\{  \omega_{k}\right\}  $ of coordinates $\left\{
q_{k}\right\}  $ and $\left\{  p_{k}\right\}  $, and masses $\left\{
m_{k}\right\}  $ (the subscript indicating the $k$th oscillator reservoir)
reads
\begin{equation}
H_{R}=\frac{1}{2}\sum\limits_{k}\left(  \frac{p_{k}^{2}}{m_{k}}+m_{k}%
\omega_{k}^{2}q_{k}^{2}\right)  \text{;} \label{2}%
\end{equation}
and the interaction Hamiltonian, linear by hypothesis and defined by the
coupling constants $C_{k}$, is given by%
\begin{equation}
H_{S/R}=q\sum\limits_{k}C_{k}q_{k}\text{.} \label{3}%
\end{equation}

The double CL model considered here consists of two quantum systems $S_{1}$
and $S_{2}$, of masses $m_{1}$ and $m_{2}$, coupled through the general form
\begin{equation}
H_{{S_{1}/}{S_{2}}}=\lambda_{11}q_{1}q_{2}+\lambda_{12}q_{1}p_{2}+\lambda
_{21}q_{2}p_{1}+\lambda_{22}p_{1}p_{2}\text{.} \label{4}%
\end{equation}
Two different situations arise, however, in the system-reservoir interactions:
$i)$ one that seems more appropriate for most physical systems, where each
oscillator is coupled to its own reservoir, and $ii)$ another, which in
practice is rather unusual, where both oscillators are coupled to a common
reservoir. The Hamiltonians governing the evolutions of the two coupled
dissipative systems are given, respectively, by
\begin{subequations}
\label{5}%
\begin{align}
H_{i}  &  =\sum_{\ell}\left(  H_{S_{\ell}}+H_{S_{\ell}/R_{\ell}}+H_{R_{\ell}%
}\right)  +H_{S_{1}/S_{2}\text{,}}\label{5a}\\
H_{ii}  &  =\sum_{\ell}\left(  H_{S_{\ell}}+H_{S_{\ell}/R}\right)
+H_{R}+H_{S_{1}/S_{2}}\text{,} \label{5b}%
\end{align}
where $\ell,\ell^{\prime},\ell^{\prime\prime}=1,2$ from here on.

In the context of a network of cavities coupled by superconducting waveguides
\cite{Raimond,Girvin,Hartmann}, distinct reservoirs must be assumed, in
general, for distinguishable cavities, even if they exhibit equal quality
factors, as long as there are no correlations whatsoever between the
reservoirs. The same applies to distinguishable trapped ions or a traveling
field reaching distinguishable optical elements. However, as demonstrated in
Ref. \cite{Mickel-AP2}, a sufficiently strong coupling between the cavities or
the trapped ions lead to a correlation between the reservoirs since, as
expected, each particular system of the network start to interact with all the
reservoirs. There are a few particular situations where a set of quantum
systems may interact with a common reservoir, such as an atomic sample or
distinct fields inside a perfect closed cavity. In the former case, different
atomic transitions couple with different reservoir modes and again, the
correlations between these reservoir modes define either a common or distinct
reservoirs, as will be demonstrated below. In the latter, it has been showed
that the proximity of the distinct field modes sets the strength of the
correlation function between the reservoir modes, which governs the emergence
of both R-DFSs \cite{Mickel-AP2}.

\subsection{\textbf{Distinct\ Reservoirs}}

Starting with the case of distinct reservoirs, the term of the Hamiltonian
accounting for the oscillators and their interactions with the reservoirs is
given by
\end{subequations}
\begin{equation}
H_{i}=\sum\limits_{\ell}\left[  \frac{p_{\ell}^{2}}{2m_{\ell}}+{V_{\ell}%
}(q_{\ell})+\sum\limits_{k}\left(  \frac{p_{\ell k}^{2}}{2m_{\ell k}}%
+\frac{m_{\ell k}}{2}\omega_{\ell k}^{2}q_{\ell k}^{2}+C_{\ell k}q_{\ell
}q_{\ell k}\right)  \right]  +H_{S_{1}/S_{2}}\text{.} \label{6}%
\end{equation}
The Lagrangian associated with Hamiltonian $H_{i}$, which defines the action
$\mathcal{S}=\int_{0}^{t}\mathcal{L}_{i}$ $d\tau$ of the Feynman-Vernon
theory, is given by%
\begin{equation}
\mathcal{L}_{i}=\sum_{\ell}\left\{  \frac{\mu_{\ell}\dot{q}_{\ell}^{2}}{2}%
-{V}_{\ell}(q_{\ell})+\sum_{k}\left[  \frac{m_{\ell k}}{2}\left(  \dot
{q}_{\ell k}^{2}-\omega_{\ell k}^{2}q_{\ell k}^{2}\right)  -C_{\ell k}q_{\ell
}q_{\ell k}\right]  \right\}  -%
\mathcal{L}%
\text{,} \label{7}%
\end{equation}
where $%
\mathcal{L}%
$, the Lagrangian associated with the interaction between the two systems,
reads
\begin{equation}%
\mathcal{L}%
=\lambda_{11}q_{1}q_{2}+\lambda_{21}\mu_{11}q_{2}\left(  \dot{q}_{1}%
-\frac{\lambda_{21}}{2}q_{2}\right)  +\lambda_{12}\mu_{22}q_{1}\left(  \dot
{q}_{2}-\frac{\lambda_{12}}{2}q_{1}\right)  +\mu_{12}\left(  \dot{q}%
_{1}-\lambda_{21}q_{2}\right)  \left(  \dot{q}_{2}-\lambda_{12}q_{1}\right)
\text{,} \label{8}%
\end{equation}
in which
\begin{equation}
\mu_{\ell\ell^{\prime}}=\frac{m_{\ell}\left(  \lambda_{22}m_{\ell^{\prime}%
}\right)  ^{1-\delta_{\ell\ell^{\prime}}}}{1-\lambda_{22}^{2}m_{1}m_{2}}
\label{8.5}%
\end{equation}
stands for reduced masses which, remarkably, arise exclusively from the
$p_{1}p_{2}$ coupling between the systems.

\subsection{\textbf{A Common\ Reservoir}}

When a common reservoir is assumed, the term of the Hamiltonian accounting for
both oscillators and their interaction with the reservoir becomes
\begin{equation}
H_{ii}=\sum\limits_{\ell}\left(  \frac{p_{\ell}^{2}}{2m_{\ell}}+{V_{\ell}%
}(q_{\ell})+\sum\limits_{k}C_{\ell k}q_{\ell}q_{k}\right)  +\frac{1}{2}%
\sum\limits_{k}\left(  \frac{p_{k}^{2}}{m_{k}}+m_{k}\omega_{k}^{2}q_{k}%
^{2}\right)  +H_{S_{1}/S_{2}}\text{.} \label{9}%
\end{equation}
The Lagrangian following from Hamiltonian $H_{ii}$, is given by%
\begin{equation}
\mathcal{L}_{ii}=\sum_{\ell}\left(  \frac{\mu_{\ell}\dot{q}_{\ell}^{2}}%
{2}-{V_{\ell}}(q_{\ell})-\sum_{k}C_{\ell k}q_{\ell}q_{k}\right)  +\sum
_{k}\frac{m_{k}}{2}\left(  \dot{q}_{k}^{2}-\omega_{k}^{2}q_{k}^{2}\right)  -%
\mathcal{L}%
\label{10}%
\end{equation}
where $%
\mathcal{L}%
$ is as defined in Eq. (\ref{8}).

Before proceeding to the calculations of the propagator for the double CL
model through the influence-functional method of Feynman and Vernon, we first
diagonalize the Hamiltonian describing both coupled systems
\begin{equation}
H_{S_{1}+S_{2}}=\sum_{\ell}\left(  \frac{p_{\ell}^{2}}{2m_{\ell}}+{V_{\ell}%
}(q_{\ell})\right)  +H_{S_{1}/S_{2}}. \label{10.5}%
\end{equation}
This diagonalization is indispensable to define the strength of the
interaction parameters $\left\{  \lambda_{\ell\ell^{\prime}}\right\}  $ which
result in positive values for the normal-mode frequencies. Otherwise, we could
have started from a positive-definite Hamiltonian with a lower bound for the
energy spectrum \cite{Mickel-AP1,Mickel,MickelRG,FLO,Amir1}.

\section{Diagonalization of the Coupled Systems}

Assuming, from here on, that the coupled systems $S_{1}$ and $S_{2}$ are
harmonic oscillators of frequencies $\omega_{1}$ and $\omega_{2}$, the
diagonalized Hamiltonian is described in terms of unitary masses and the
normal-mode coordinates $Q_{\ell}$ and $P_{\ell}$, defined in Eqs. (\ref{A8}),
as
\begin{equation}
H_{S_{1}+S_{2}}=\frac{1}{2}\sum\limits_{\ell}\left(  P_{\ell}^{2}+\Omega
_{\ell}^{2}Q_{\ell}^{2}\right)  \text{.}\label{11}%
\end{equation}
The original masses $m_{\ell}$ have been absorbed by the normal-mode
frequencies $\Omega_{\ell}^{2}$, obtained in Eq. (\ref{A3}), which account for
the effective interactions $g_{\ell}$, given in Eq. (\ref{A2}), apart from the
natural frequencies $\omega_{1}$ and $\omega_{2}$. With the condition that the
normal-mode frequencies $\Omega_{\ell}$ assume positive values, it is
straightforward to show that the relations%
\begin{subequations}
\begin{align}
\left(  \frac{\omega_{1}-\omega_{2}}{2}\right)  ^{2}+\left\vert g_{2}%
\right\vert ^{2} &  \geq\left\vert g_{1}\right\vert ^{2}\left(  \frac
{\omega_{1}-\omega_{2}}{\omega_{1}+\omega_{2}}\right)  ^{2}\text{,}%
\label{12b}\\
\left(  \left\vert g_{2}\right\vert ^{2}-\left\vert g_{1}\right\vert
^{2}\right)  ^{2}+\left(  \omega_{1}\omega_{2}\right)  ^{2} &  \geq2\omega
_{1}\omega_{2}\left(  \left\vert g_{1}\right\vert ^{2}+\left\vert
g_{2}\right\vert ^{2}\right)  \text{,}\label{12c}%
\end{align}
must be satisfied.

The coordinates $Q_{\ell}$ and $P_{\ell}$, associated with the normal-mode
frequencies, follow from the previous generalized coordinates $q_{\ell}$ and
$p_{\ell}$, as described by the transformation (\ref{A6}). With these
normal-mode frequencies and coordinates, the full Hamiltonians $H_{i}$ and
$H_{ii}$ becomes:
\end{subequations}
\begin{subequations}
\begin{align}
H_{i}  &  =\frac{1}{2}\sum\limits_{\ell}\left(  P_{\ell}^{2}+\Omega_{\ell}%
^{2}Q_{\ell}^{2}+2\sqrt{\frac{2\hbar}{m_{\ell}\omega_{\ell}}}\sum
\limits_{\ell^{\prime}}\left[  \mathrm{\operatorname{Re}}\left(  c_{\ell
\ell^{\prime}}Q_{\ell^{\prime}}\right)  +\operatorname{Im}\left(  d_{\ell
\ell^{\prime}}P_{\ell^{\prime}}\right)  \right]  \sum\limits_{k}C_{\ell
k}q_{k}\right) \nonumber\\
&  +\frac{1}{2}\sum\limits_{k}\left(  \frac{p_{k}^{2}}{m_{k}}+m_{k}\omega
_{k}^{2}q_{k}^{2}\right)  \text{,}\label{17a}\\
H_{ii}  &  =\frac{1}{2}\sum\limits_{\ell}\left[  P_{\ell}^{2}+\Omega_{\ell
}^{2}Q_{\ell}^{2}+\sum\limits_{k}\left(  \frac{p_{\ell k}^{2}}{m_{\ell k}%
}+m_{\ell k}\omega_{\ell k}^{2}q_{\ell k}^{2}\right.  \right. \nonumber\\
&  \left.  \left.  +2\sqrt{\frac{2\hbar}{m_{\ell}\omega_{\ell}}}%
\sum\limits_{\ell^{\prime}}\left[  \mathrm{\operatorname{Re}}\left(
c_{\ell\ell^{\prime}}Q_{\ell^{\prime}}\right)  +\operatorname{Im}\left(
d_{\ell\ell^{\prime}}P_{\ell^{\prime}}\right)  \right]  C_{\ell k}q_{\ell
k}\right)  \right]  \text{,} \label{17b}%
\end{align}
where we have used the inverse of the transformation Eq. (\ref{A6}), given by
\end{subequations}
\begin{subequations}
\label{19}%
\begin{align}
q_{\ell}  &  =\sqrt{\frac{2\hbar}{m_{\ell}\omega_{\ell}}}\sum\limits_{\ell
^{\prime}}\left[  \mathrm{\operatorname{Re}}\left(  c_{\ell\ell^{\prime}%
}Q_{\ell^{\prime}}\right)  +\operatorname{Im}\left(  d_{\ell\ell^{\prime}%
}P_{\ell^{\prime}}\right)  \right]  \text{,}\label{19a}\\
p_{\ell}  &  =\sqrt{2\hbar m_{\ell}\omega_{\ell}}\sum\limits_{\ell^{\prime}%
}\left[  \mathrm{\operatorname{Re}}\left(  d_{\ell\ell^{\prime}}%
P_{\ell^{\prime}}\right)  -\operatorname{Im}\left(  c_{\ell\ell^{\prime}%
}Q_{\ell^{\prime}}\right)  \right]  \text{,} \label{19b}%
\end{align}
with the coefficients
\end{subequations}
\begin{subequations}
\label{20}%
\begin{align}
c_{\ell\ell^{\prime}}  &  =\mathcal{N}_{\ell^{\prime}}\sqrt{\frac{\Omega
_{\ell^{\prime}}}{2\hslash}}\left[  -\Delta_{1\ell}\left(  \Omega
_{\ell^{\prime}}\right)  +\Delta_{2\ell}\left(  \Omega_{\ell^{\prime}}\right)
\right]  \operatorname{e}^{i\phi_{\ell}}\text{,}\label{20a}\\
d_{\ell\ell^{\prime}}  &  =\mathcal{N}_{\ell^{\prime}}\sqrt{\frac{1}%
{2\hslash\Omega_{\ell^{\prime}}}}\left[  \Delta_{1\ell}\left(  \Omega
_{\ell^{\prime}}\right)  +\Delta_{2\ell}\left(  \Omega_{\ell^{\prime}}\right)
\right]  \operatorname{e}^{i\phi_{\ell}}\text{,} \label{20b}%
\end{align}
and the functions $\Delta_{\ell\ell^{\prime}}\left(  \Omega_{\ell^{\prime}%
}\right)  $ given by Eq. (\ref{A5b}).

Therefore, through the diagonalized form of the Hamiltonian $H_{S_{1}+S_{2}}$,
we get two independent and generalized CL Hamiltonians in terms of normal
coordinates $Q_{\ell}$ and $P_{\ell}$. The Hamiltonians (\ref{17a}) and
(\ref{17b}) describe two independent harmonic oscillators, both interacting
with a common reservoir in the latter case and each one interacting with its
own reservoir in the former. By a generalized CL Hamiltonian we mean that the
system-reservoir coupling exhibits, due to the general form of the interaction
between the two oscillators in Eq. (\ref{5}), a momentum-position term apart
from the usual position-position one. Thus, differently from the CL model,
where a momentum-position coupling is easily handled through a coordinate
transformation \cite{CL}, in the present double CL model this
momentum-position coupling hampers the application of the influence-functional
method, since the derived Lagrangian involves tricky reservoir-reservoir
interaction terms. Therefore, in spite of the diagonalization of the
interaction described by Hamiltonian $H_{S_{1}+S_{2}}$, it is preferable to
approach the problem, for both distinct reservoirs and a common one, through
the Lagrangians in Eqs. (\ref{7}) and (\ref{10}), respectively. As
demonstrated below, in the former case we end up with a product of two
influence functionals, identical to that of the CL model, whereas for a common
reservoir the influence functional does not factorize into individual functionals.

\section{Feynman-Vernon theory}

To obtain the reduced master equation describing the time evolution of the
coupled systems $S_{1}$ and $S_{2}$, we proceed from the integral form of the
density operator of the whole system at a time $t$, given by $\rho
(t)=U(t)\rho(0)U^{\dagger}(t)$, where the evolution operator $U(t)$ follows
from Hamiltonians $H_{i}$ and $H_{ii}$. Next, we assign the coordinates
$x_{\ell}$ and $y_{\ell}$ to system $S_{\ell}$ and the $N$-component vectors
$\mathbf{X}_{\ell}=(X_{\ell1},\ldots,X_{\ell N})$ and $\mathbf{Y}_{\ell
}=(Y_{\ell1},\ldots,Y_{\ell N})$ to reservoirs $R_{\ell}$. In the case of a
common reservoir, we only have to disregard one of the reservoirs $R_{\ell}$,
in the expression derived for the case of distinct reservoirs, to obtain the
associated influence functional. Using the notation $\left\{  x_{\ell
}\right\}  =x_{1}$,$x_{2}$ and $\left\{  \operatorname{d}x_{\ell}\right\}
=\operatorname{d}x_{1}\operatorname{d}x_{2}$, and the same form for other
variables, we obtain the matrix element in the coordinate representation
\end{subequations}
\begin{align}
&  \left\langle \left\{  x_{\ell}\right\}  ,\left\{  \mathbf{X}_{\ell
}\right\}  \left\vert \rho(t)\right\vert \left\{  y_{\ell}\right\}  ,\left\{
\mathbf{Y}_{\ell}\right\}  \right\rangle \nonumber\\
&  =\int\operatorname{d}\left\{  x_{\ell}^{\prime}\right\}  \operatorname{d}%
\left\{  y_{\ell}^{\prime}\right\}  \operatorname{d}\left\{  \mathbf{X}_{\ell
}^{\prime}\right\}  \operatorname{d}\left\{  \mathbf{Y}_{\ell}^{\prime
}\right\}  \text{ }\left\langle \left\{  x_{\ell}^{\prime}\right\}  ,\left\{
\mathbf{X}_{\ell}^{\prime}\right\}  \left\vert \rho(0)\right\vert \left\{
y_{\ell}^{\prime}\right\}  ,\left\{  \mathbf{Y}_{\ell}^{\prime}\right\}
\right\rangle \nonumber\\
&  \times K\left(  \left\{  x_{\ell}\right\}  ,\left\{  \mathbf{X}_{\ell
}\right\}  ,t;\left\{  x_{\ell}^{\prime}\right\}  ,\left\{  \mathbf{X}_{\ell
}^{\prime}\right\}  ,0\right)  K^{\ast}\left(  \left\{  y_{\ell}\right\}
,\left\{  \mathbf{Y}_{\ell}\right\}  ,t;\left\{  y_{\ell}^{\prime}\right\}
,\left\{  \mathbf{Y}_{\ell}^{\prime}\right\}  ,0\right)  \text{,} \label{21}%
\end{align}
where functional integrations are evaluated over paths $x_{\ell}(t^{\prime})$,
$y_{\ell}(t^{\prime})$, $\mathbf{X}_{\ell}(t^{\prime})$ and $\mathbf{Y}_{\ell
}(t^{\prime})$, with endpoints\ $x_{\ell}(t)=x_{\ell}$, $x_{\ell}(0)=x_{\ell
}^{\prime}$, $y_{\ell}(t)=y_{\ell}$, $y_{\ell}(0)=y_{\ell}^{\prime}$,
$\mathbf{X}_{\ell}(t)=\mathbf{X}_{\ell}$, $\mathbf{X}_{\ell}(0)=\mathbf{X}%
_{\ell}^{\prime}$, $\mathbf{Y}_{\ell}(t)=\mathbf{Y}_{\ell}$, and
$\mathbf{Y}_{\ell}(0)=\mathbf{Y}_{\ell}^{\prime}$. The propagator $K$ is given
by
\begin{subequations}
\label{22}%
\begin{equation}
K\left(  \left\{  x_{\ell}\right\}  ,\left\{  \mathbf{X}_{\ell}\right\}
,t;\left\{  x_{\ell}^{\prime}\right\}  ,\left\{  \mathbf{X}_{\ell}^{\prime
}\right\}  ,0\right)  =%
{\displaystyle\int}
\operatorname{D}\left\{  x_{\ell}\right\}  \operatorname{D}\left\{
\mathbf{X}_{\ell}\right\}  \text{ }\exp\left(  \frac{i}{\hslash}%
\mathcal{S}\left[  \left\{  x_{\ell}\right\}  ,\left\{  \mathbf{X}_{\ell
}\right\}  \right]  \right)  \text{.} \label{22b}%
\end{equation}

The action $\mathcal{S}$ follows from the Lagrangian $\mathcal{L}$
$=\mathcal{L}_{i}$ or $\mathcal{L}_{ii}$, as $\mathcal{S}=%
{\displaystyle\int\nolimits_{0}^{t}}
\mathcal{L}\operatorname{d}t^{\prime}$. Tracing out the reservoir coordinates,
we obtain the reduced density operator describing the coupled systems $S_{1}$
and $S_{2}$ under the influence of their respective reservoir, given by%
\end{subequations}
\begin{equation}
\widetilde{\rho}\left(  \left\{  x_{\ell}\right\}  ,\left\{  y_{\ell}\right\}
,t\right)  =\int\operatorname{d}\left\{  x_{\ell}^{\prime}\right\}
\operatorname{d}\left\{  y_{\ell}^{\prime}\right\}  \text{ }\operatorname{J}%
\left(  \left\{  x_{\ell}\right\}  ,\left\{  y_{\ell}\right\}  ,t;\left\{
x_{\ell}^{\prime}\right\}  ,\left\{  y_{\ell}^{\prime}\right\}  ,0\right)
\widetilde{\rho}\left(  \left\{  x_{\ell}^{\prime}\right\}  ,\left\{  y_{\ell
}^{\prime}\right\}  ,0\right)  \text{,} \label{23}%
\end{equation}
where we have assumed that the system-reservoir coupling is turned on
suddenly, such that the total density operator is initially given by
$\rho(0)=\widetilde{\rho}(0)\rho_{R_{1}+R_{2}}(0)$. The propagator for the
density operator turns out to be%
\begin{align}
&  \operatorname{J}\left(  \left\{  x_{\ell}\right\}  ,\left\{  y_{\ell
}\right\}  ,t;\left\{  x_{\ell}^{\prime}\right\}  ,\left\{  y_{\ell}^{\prime
}\right\}  ,0\right) \nonumber\\
&  =%
{\displaystyle\int}
\operatorname{D}\left\{  x_{\ell}\right\}  \operatorname{D}\left\{  y_{\ell
}\right\}  \text{ }%
\mathcal{F}%
\left[  \left\{  x_{\ell}\right\}  ,\left\{  y_{\ell}\right\}  \right]
\nonumber\\
&  \times\exp\left\{  \frac{i}{\hbar}\left[  \sum_{\ell}\left(  \mathcal{S}%
_{S_{\ell}}\left[  \left\{  x_{\ell}\right\}  \right]  -\mathcal{S}_{S_{\ell}%
}\left[  \left\{  y_{\ell}\right\}  \right]  \right)  +\mathcal{S}%
_{S_{1}/S_{2}}\left[  \left\{  x_{\ell}\right\}  \right]  -\mathcal{S}%
_{S_{1}/S_{2}}\left[  \left\{  y_{\ell}\right\}  \right]  \right]  \right\}
\text{.} \label{24}%
\end{align}
Although expression (\ref{23}) applies to both cases, $i)$ and $ii)$, the
computation of the influence functional $%
\mathcal{F}%
\left[  \left\{  x_{\ell}\right\}  ,\left\{  y_{\ell}\right\}  \right]  $
\cite{22,23} representing the effects of the reservoirs on the systems is
completely different for the two cases.

\subsection{\textbf{Distinct\ Reservoirs}}

When considering distinct reservoirs, the influence functional in Eq.
(\ref{24}) simply factorizes as
\begin{equation}%
\mathcal{F}%
\left[  \left\{  x_{\ell}\right\}  ,\left\{  y_{\ell}\right\}  \right]
=\prod_{\ell}%
\mathcal{F}%
\left[  x_{\ell},y_{\ell}\right]  \text{,} \label{25}%
\end{equation}
and the component arising from the interaction of system $S_{\ell}$ with
reservoir $R_{\ell}$, is written as%
\begin{align}%
\mathcal{F}%
\left[  x_{\ell},y_{\ell}\right]   &  =\int\operatorname{d}\mathbf{X}_{\ell
}\operatorname{d}\mathbf{X}_{\ell}^{\prime}\operatorname{d}\mathbf{Y}_{\ell
}^{\prime}\text{ }\rho_{R_{\ell}}\left(  \mathbf{X}_{\ell}^{\prime}%
,\mathbf{Y}_{\ell}^{\prime},0\right)
{\displaystyle\int}
\operatorname{D}\mathbf{X}_{\ell}\operatorname{D}\mathbf{Y}_{\ell}\text{
}\nonumber\\
&  \times\exp\left\{  \frac{i}{\hbar}\left(  \mathcal{S}_{R_{\ell}}\left[
\mathbf{X}_{\ell}\right]  -\mathcal{S}_{R_{\ell}}\left[  \mathbf{Y}_{\ell
}\right]  +\mathcal{S}_{S_{\ell}/R_{\ell}}\left[  x_{\ell},\mathbf{X}_{\ell
}\right]  -\mathcal{S}_{S_{\ell}/R_{\ell}}\left[  y_{\ell},\mathbf{Y}_{\ell
}\right]  \right)  \right\}  \label{26}%
\end{align}
\qquad\qquad Evidently, when disregarding the interaction between the systems,
described by Hamiltonian $H_{{S_{1}/}{S_{2}}}$ in Eq. (\ref{4}), the
propagator $\operatorname{J}$ in Eq. (\ref{24}) reduces to a product of
propagators identical to that obtained in the CL model. However, the influence
functional factorizes independently of any requirement for non-interacting
systems, since it only takes into account the interaction between the systems
and their respective reservoirs. Therefore, the influence functional
(\ref{25}) is obtained directly as a product of the functional obtained from
the CL model. All that remains to be done is to obtain the propagator
(\ref{24}) by considering the actions coming from the interacting systems.

Thus, from the CL model we obtain directly the form%
\begin{align}%
\mathcal{F}%
\left[  x_{\ell},y_{\ell}\right]   &  =\exp\left\{  -\frac{1}{\hbar}\int
_{0}^{t}\operatorname{d}\tau\int_{0}^{\tau}\operatorname{d}t^{\prime}\text{
}\alpha_{\ell\operatorname*{R}}(\tau-t^{\prime})\left[  x_{\ell}(\tau
)-y_{\ell}(\tau)\right]  \left[  x_{\ell}(t^{\prime})-y_{\ell}(t^{\prime
})\right]  \right\} \nonumber\\
&  \times\exp\left\{  -\frac{i}{\hbar}\int_{0}^{t}\operatorname{d}\tau\int
_{0}^{\tau}\operatorname{d}t^{\prime}\text{ }\alpha_{\ell\operatorname{I}%
}(\tau-t^{\prime})\left[  x_{\ell}(\tau)-y_{\ell}(\tau)\right]  \left[
x_{\ell}(t^{\prime})+y_{\ell}(t^{\prime})\right]  \right\}  \text{,}
\label{27}%
\end{align}
with the real and imaginary parts of a function $\alpha\left(  t\right)  $
given by
\begin{subequations}
\label{28}%
\begin{align}
\alpha_{\ell\operatorname{R}}(t)  &  =\sum_{k}\frac{C_{\ell k}^{2}}{2m_{\ell
k}\omega_{\ell k}}\coth\left(  \omega_{\ell k}\hbar\beta_{\ell}/2\right)
\cos\left[  \omega_{\ell k}(t)\right]  \text{,}\label{28a}\\
\alpha_{\ell\operatorname{I}}(t)  &  =-\sum_{k}\frac{C_{\ell k}^{2}}{2m_{\ell
k}\omega_{\ell k}}\sin\left[  \omega_{\ell k}(t)\right]  \text{.} \label{28b}%
\end{align}

Assuming that the reservoir modes are sufficiently closely spaced to allow a
continuum summation, we define the spectral functions \cite{CL}%
\end{subequations}
\begin{equation}
\chi_{\ell}(\omega)=\pi\sum_{k}\frac{C_{\ell k}^{2}}{2m_{\ell k}\omega_{\ell
k}}\delta(\omega-\omega_{\ell k})\text{.}\label{29}%
\end{equation}
The introduction of a frequency cutoff $\Omega_{\ell}^{C}$ considerably higher
than the characteristic frequencies of the problem, together with the
assumption of an Ohmic reservoir, where the distributions $\chi_{\ell}%
(\omega)$ are defined by the damping constants $\eta_{\ell}$, such that
\begin{equation}
\chi_{\ell}(\omega)=\left\{
\begin{array}
[c]{cc}%
\eta_{\ell}\omega\text{,} & \omega<\Omega_{\ell}^{C}\text{,}\\
0\text{,} & \omega>\Omega_{\ell}^{C}\text{,}%
\end{array}
\right.  \label{30}%
\end{equation}
enable us to rewrite Eq. (\ref{28}) as
\begin{subequations}
\label{31}%
\begin{align}
\alpha_{\ell\operatorname{R}}(t) &  =\frac{1}{\pi}\int_{0}^{\Omega_{\ell}^{C}%
}\operatorname{d}\omega_{\ell}\text{ }\eta_{\ell}\omega_{\ell}\coth\left(
\omega_{\ell}\hbar\beta/2\right)  \cos\left[  \omega_{\ell}(t)\right]
\text{,}\label{31a}\\
\alpha_{\ell\operatorname{I}}(t) &  =-\frac{1}{\pi}\int_{0}^{\Omega_{\ell}%
^{C}}\operatorname{d}\omega_{\ell}\text{ }\eta_{\ell}\omega_{\ell}\sin\left[
\omega_{\ell}(t)\right]  \nonumber\\
&  =\frac{\eta_{\ell}}{\pi t}\left(  \Omega_{\ell}^{C}\cos\left[  \Omega
_{\ell}^{C}(t)\right]  -\frac{\sin\left[  \Omega_{\ell}^{C}(t)\right]  }%
{t}\right)  \text{.}\label{31b}%
\end{align}
As we are interested in times much longer than the typical value
$1/\Omega_{\ell}^{C}$, it follows that $\sin\left[  \Omega_{\ell}^{C}\left(
t\right)  \right]  /\pi t\approx\delta\left(  t\right)  $ and, consequently
\end{subequations}
\begin{subequations}
\label{32}%
\begin{align}
\int_{0}^{t}\operatorname{d}\tau\text{ }\frac{\sin\left[  \Omega_{\ell}%
^{C}(\tau)\right]  }{\pi\tau}\left[  x_{\ell}(\tau)-y_{\ell}(\tau)\right]   &
\approx0\text{,}\label{32a}\\
\int_{0}^{\tau}\operatorname{d}t\text{ }\left[  \dot{x}_{\ell}(t)+\dot
{y}_{\ell}(t)\right]  \frac{\sin\left[  \Omega_{\ell}^{C}(\tau-t)\right]
}{\pi\left(  \tau-t\right)  } &  \approx\frac{\dot{x}_{\ell}(\tau)+\dot
{y}_{\ell}(\tau)}{2}\text{.}\label{32b}%
\end{align}
With these approximations, the propagator becomes%
\end{subequations}
\begin{align}
&  \operatorname{J}\left(  \left\{  x_{\ell}\right\}  ,\left\{  y_{\ell
}\right\}  ,t;\left\{  x_{\ell}^{\prime}\right\}  ,\left\{  y_{\ell}^{\prime
}\right\}  ,0\right)  \nonumber\\
&  =\exp\left[  -i\sum_{\ell}\frac{\gamma_{\ell}\mu_{\ell\ell}}{2\hbar}\left(
x_{\ell}^{2}-x_{\ell}^{\prime2}-y_{\ell}^{2}+y_{\ell}^{\prime2}\right)
\right]
{\displaystyle\int}
\operatorname{D}\left\{  x_{\ell}\right\}  \operatorname{D}\left\{  y_{\ell
}\right\}  \nonumber\\
&  \times\exp\left\{  \frac{i}{\hbar}\left[  \sum_{\ell}\left(  \widetilde
{\mathcal{S}}_{S_{\ell}}[x_{\ell}]-\widetilde{\mathcal{S}}_{S_{\ell}}[y_{\ell
}]\right)  +\mathcal{S}_{S_{1}/S_{2}}\left[  \left\{  x_{\ell}\right\}
\right]  -\mathcal{S}_{S_{1}/S_{2}}\left[  \left\{  y_{\ell}\right\}  \right]
\right]  \right\}  \nonumber\\
&  \times\exp\left\{  -\sum_{\ell}\frac{\gamma_{\ell}\mu_{\ell\ell}}{\hbar
}\left[  \frac{2}{\pi}\int_{0}^{t}\operatorname{d}\tau\int_{0}^{\tau
}\operatorname{d}t^{\prime}\text{ }\left[  x_{\ell}(\tau)-y_{\ell}%
(\tau)\right]  \left[  x_{\ell}(t^{\prime})-y_{\ell}(t^{\prime})\right]
\right.  \right.  \nonumber\\
&  \times\int_{0}^{\Omega_{\ell}^{C}}\operatorname{d}\omega_{\ell}\text{
}\omega_{\ell}\coth\left(  \omega_{\ell}\hbar\beta_{\ell}/2\right)
\cos\left[  \omega_{\ell}(\tau-t^{\prime})\right]  \nonumber\\
&  \left.  \left.  -i\int_{0}^{t}\operatorname{d}\tau\text{ }\left[  x_{\ell
}(\tau)\dot{y}_{\ell}(\tau)-y_{\ell}(\tau)\dot{x}_{\ell}(\tau)\right]
\right]  \right\}  \text{,}\label{33}%
\end{align}
where we have defined the\ relaxation constant $\gamma_{\ell}=\eta_{\ell}%
/2\mu_{\ell\ell}$ for the two system-reservoir couplings and the renormalized
actions%
\begin{subequations}
\begin{align}
\widetilde{\mathcal{S}}_{S_{\ell}}[x_{\ell}] &  =\int_{0}^{t}\operatorname{d}%
\tau\text{ }\mathcal{L}_{i}(x_{\ell},\dot{x}_{\ell},t)\nonumber\\
&  =\int_{0}^{t}\operatorname{d}\tau\text{ }\left[  \frac{\mu_{\ell\ell}}%
{2}\dot{x}_{\ell}^{2}-\widetilde{{V}}{_{\ell}(x_{\ell})}\right]
\text{,}\label{34a}\\
\mathcal{S}_{S_{1}/S_{2}}\left[  \left\{  x_{\ell}\right\}  \right]   &
=\int_{0}^{t}\operatorname{d}\tau\text{ }L_{12}\left(  \left\{  x_{\ell
}\right\}  ,\left\{  \dot{x}_{\ell}\right\}  ,t^{\prime}\right)  \nonumber\\
&  =-\int_{0}^{t}\operatorname{d}\tau\text{ }\left[  \lambda_{11}x_{1}%
x_{2}+\mu_{12}\left(  \dot{x}_{1}-\lambda_{21}x_{2}\right)  \left(  \dot
{x}_{2}-\lambda_{12}x_{1}\right)  \right.  \nonumber\\
&  \left.  +\lambda_{12}\mu_{22}x_{1}\left(  \dot{x}_{2}-\frac{\lambda_{12}%
}{2}x_{1}\right)  +\lambda_{21}\mu_{11}x_{2}\left(  \dot{x}_{1}-\frac
{\lambda_{21}}{2}x_{2}\right)  \right]  \text{,}\label{34b}%
\end{align}
with similar expressions for the variables $y_{\ell}$. The renormalized
potentials $\widetilde{{V}}{_{\ell}(x_{\ell})}=m_{\ell}\left[  \omega_{\ell
}^{2}-\left(  \Delta\omega_{\ell}\right)  ^{2}\right]  x_{\ell}^{2}/2$, follow
from the system-reservoir couplings which induce the shifts $\left(
\Delta\omega_{\ell}\right)  ^{2}=2\eta_{\ell}\Omega_{\ell}^{C}/\pi m_{\ell}$.
Thus, we can define a renormalized frequency, given by $\widetilde{\omega
}_{\ell}^{2}=\omega_{\ell}^{2}-\left(  \Delta\omega_{\ell}\right)  ^{2}%
$\cite{CL}.

From the result that the functional integral for an infinitesimal time
evolution can be approximated by \cite{22},
\end{subequations}
\begin{equation}
\int\operatorname{D}q\exp\left(  \frac{i}{\hbar}\mathcal{S}\right)
\approx\frac{1}{N}\exp\left(  \frac{i}{\hbar}\mathcal{S}\right)  \text{,}
\label{35}%
\end{equation}
we consider the evolution of the reduced density operator $\widetilde{\rho}$
on the infinitesimal time interval between $t$ and $t+\varepsilon$
($\varepsilon\rightarrow0$), proceeding from Eq. (\ref{22}), to obtain
\begin{align}
&  \widetilde{\rho}\left(  \left\{  x_{\ell}\right\}  ,\left\{  y_{\ell
}\right\}  ,t+\varepsilon\right) \nonumber\\
&  =\int\operatorname{d}\left\{  x_{\ell}^{\prime}\right\}  \operatorname{d}%
\left\{  y_{\ell}^{\prime}\right\}  \text{ }\operatorname{J}\left(  \left\{
x_{\ell}\right\}  ,\left\{  y_{\ell}\right\}  ,t+\varepsilon;\left\{  x_{\ell
}^{\prime}\right\}  ,\left\{  y_{\ell}^{\prime}\right\}  ,t\right)
\widetilde{\rho}\left(  \left\{  x_{\ell}^{\prime}\right\}  ,\left\{  y_{\ell
}^{\prime}\right\}  ,t\right)  \text{.} \label{36}%
\end{align}
We also assume the high-temperature limit $k_{B}T_{\ell}\gg\hbar\omega_{\ell}$
(for both reservoir frequencies $\omega_{\ell}\ll\Omega_{\ell}^{C}$), which
allows analytical solutions of the integrals (in the variable $\omega_{\ell}$)
when defining the propagator $\operatorname{J}$ and using, for any function
$\operatorname{F}$, the approximations $\dot{x}_{\ell}\approx\beta_{1\ell
}/\varepsilon$, $\dot{y}_{\ell}\approx\beta_{2\ell}/\varepsilon$, and%

\begin{equation}
\int_{t}^{t+\varepsilon}\operatorname{d}\tau\operatorname{F}\left[  x_{\ell
}(\tau)\right]  \approx\varepsilon\operatorname{F}\left[  \frac{x_{\ell
}+x_{\ell}^{\prime}}{2}\right]  \text{,}%
\end{equation}
where we have defined the variables $\beta_{1\ell}=x_{\ell}-x_{\ell}^{\prime}$
and $\beta_{2\ell}=y_{\ell}-y_{\ell}^{\prime}$. With the above approximations,
we obtain from Eq. (\ref{33}) the propagator%
\begin{align}
&  \operatorname{J}\left(  \left\{  x_{\ell}\right\}  ,\left\{  y_{\ell
}\right\}  ,t+\varepsilon;\left\{  x_{\ell}-\beta_{1\ell}\right\}  ,\left\{
y_{\ell}-\beta_{2\ell}\right\}  ,t\right) \nonumber\\
&  =\mathcal{N}\exp\left\{  \frac{i}{\hbar}\left[  \frac{\mu_{12}}%
{\varepsilon}\sum_{\ell}\left(  -1\right)  ^{\ell}\beta_{\ell1}\beta_{\ell
2}+\varepsilon\left(  \lambda_{11}+\lambda_{12}\lambda_{21}\mu_{12}\right)
\left[  \prod_{\ell}\left(  y_{\ell}-\frac{\beta_{2\ell}}{2}\right)
-\prod_{\ell}\left(  x_{\ell}-\frac{\beta_{1\ell}}{2}\right)  \right]
\right.  \right. \nonumber\\
&  +\sum_{\ell,\neq\ell^{\prime}}\lambda_{\ell\ell^{\prime}}\left\{
\sum_{\ell^{\prime\prime}}\left(  -1\right)  ^{\ell^{\prime}+\ell
^{\prime\prime}}\mu_{\ell^{\prime\prime}\ell^{\prime}}\left[  \beta
_{2\ell^{\prime\prime}}\left(  y_{\ell}-\frac{\beta_{2\ell}}{2}\right)
-\beta_{1\ell^{\prime\prime}}\left(  x_{\ell}-\frac{\beta_{1\ell}}{2}\right)
\right]  \right. \nonumber\\
&  \left.  \left.  \left.  +\frac{\varepsilon}{2}\lambda_{\ell\ell^{\prime}%
}\mu_{\ell^{\prime}\ell^{\prime}}\left[  \left(  x_{\ell}-\frac{\beta_{1\ell}%
}{2}\right)  ^{2}+\left(  y_{\ell}-\frac{\beta_{2\ell}}{2}\right)
^{2}\right]  \right\}  \right]  \right\} \nonumber\\
&  \times\prod_{\ell}\exp\left\{  \frac{i\mu_{\ell\ell}}{\hbar}\left[
\frac{1}{2}\left(  \beta_{1\ell}^{2}-\beta_{2\ell}^{2}\right)  \left(
\gamma_{\ell}+\frac{1}{\varepsilon}\right)  -\gamma_{\ell}\left(  x_{\ell
}-y_{\ell}\right)  \left(  \beta_{1\ell}+\beta_{2\ell}\right)  \right]
\right. \nonumber\\
&  \left.  -\frac{i\varepsilon}{\hbar}\left[  \widetilde{V}_{\ell}\left(
x_{\ell}-\frac{\beta_{1\ell}}{2}\right)  -\widetilde{V}_{\ell}\left(  y_{\ell
}-\frac{\beta_{2\ell}}{2}\right)  \right]  -\frac{2\mu_{\ell\ell}\gamma_{\ell
}k_{B}T_{\ell}}{\hbar^{2}}\varepsilon\left[  \left(  x_{_{\ell}}-\frac
{\beta_{1\ell}}{2}\right)  -\left(  y_{_{\ell}}-\frac{\beta_{2\ell}}%
{2}\right)  \right]  ^{2}\right\}  \label{38}%
\end{align}
where $\mathcal{N}$ is a normalization factor. Since the fast-oscillating
terms in the integrals in Eq. (\ref{35}) contribute only for $\beta_{\ell
k}\approx\sqrt{\varepsilon\hslash/M}$, we expand both sides of Eq. (\ref{35})
up to terms $\mathcal{O}(\varepsilon)$ \cite{CL}. Proceeding to a further
change of variables: $\beta_{\ell\ell^{\prime}}^{\prime}=\beta_{\ell
\ell^{\prime}}+\left(  -1\right)  ^{\ell}\gamma_{\ell^{\prime}}(x_{\ell
^{\prime}}-y_{\ell^{\prime}})\varepsilon$, keeping again terms up to
$\mathcal{O}(\varepsilon)$, we obtain for the zeroth order term, the
normalization factor $\mathcal{N}=\left(  2\pi\hslash\varepsilon\right)
^{2}/\left(  \mu_{12}^{2}-\mu_{1}\mu_{2}\right)  $, and for the first order
term the desired equation of motion \cite{CL}%

\begin{align}
\frac{\partial\widetilde{\rho}}{\partial t}  &  =-\sum_{\ell}\left\{
-i\frac{\hbar}{2m_{\ell}}\left(  \frac{\partial^{2}}{\partial x_{\ell}^{2}%
}-\frac{\partial^{2}}{\partial y_{\ell}^{2}}\right)  +\widetilde{\gamma}%
_{\ell}\left(  x_{\ell}-y_{\ell}\right)  \left(  \frac{\partial}{\partial
x_{\ell}}-\frac{\partial}{\partial y_{\ell}}\right)  \right. \nonumber\\
&  +i\frac{m_{\ell}\widetilde{\omega}_{\ell}^{2}}{2\hbar}\left(  x_{\ell}%
^{2}-y_{\ell}^{2}\right)  +\frac{2m_{\ell}\widetilde{\gamma}_{\ell}%
k_{B}T_{\ell}}{\hbar^{2}}\left(  x_{\ell}-y_{\ell}\right)  ^{2}\nonumber\\
&  +\sum_{\ell^{\prime}\left(  \neq\ell\right)  }\left[  -i\frac{\hbar
\lambda_{22}}{2}\left(  \frac{\partial^{2}}{\partial x_{\ell}\partial
x_{\ell^{\prime}}}-\frac{\partial^{2}}{\partial y_{\ell}\partial
y_{\ell^{\prime}}}\right)  +\lambda_{22}m_{\ell^{\prime}}\widetilde{\gamma
}_{\ell^{\prime}}\left(  x_{\ell^{\prime}}-y_{\ell^{\prime}}\right)  \left(
\frac{\partial}{\partial x_{\ell}}-\frac{\partial}{\partial y_{\ell}}\right)
\right. \nonumber\\
&  \left.  \left.  +\lambda_{\ell^{\prime}\ell}\left(  x_{\ell^{\prime}}%
\frac{\partial}{\partial x_{\ell}}+y_{\ell^{\prime}}\frac{\partial}{\partial
y_{\ell}}\right)  +i\frac{\lambda_{\ell^{\prime}\ell}m_{\ell}\widetilde
{\gamma}_{\ell}}{\hbar}\left(  x_{\ell}-y_{\ell}\right)  \left(
x_{\ell^{\prime}}+y_{\ell^{\prime}}\right)  +i\frac{\lambda_{11}}{2\hbar
}\left(  x_{\ell}x_{\ell^{\prime}}-y_{\ell}y_{\ell^{\prime}}\right)  \right]
\right\}  \widetilde{\rho} \label{39}%
\end{align}
where we have defined the effective damping rates
\begin{equation}
\widetilde{\gamma}_{\ell}=\frac{\gamma_{\ell}}{1-\lambda_{22}^{2}m_{1}m_{2}%
}\text{,} \label{40}%
\end{equation}
which increase with increasing coupling strength $\lambda_{22}$. The operator
equation associated with the above coordinate representation turns out to be
\begin{align}
\frac{\partial\widetilde{\rho}}{\partial t}  &  =-\frac{i}{\hbar}\left[
\mathbf{H}_{i},\widetilde{\rho}\right]  +\sum_{\ell}\widetilde{\gamma}_{\ell
}\left[  \left(  -\frac{i}{\hbar}\left[  x_{\ell},\left\{  p_{\ell}%
,\widetilde{\rho}\right\}  \right]  -\frac{2m_{\ell}k_{B}T_{\ell}}{\hbar^{2}%
}\left[  x_{\ell},\left[  x_{\ell},\widetilde{\rho}\right]  \right]  \right)
\right. \label{40.5}\\
&  \left.  -i\frac{m_{\ell}}{\hbar}\sum_{\ell^{\prime}\left(  \neq\ell\right)
}\left(  \lambda_{22}\left[  x_{\ell},\left\{  p_{\ell^{\prime}}%
,\widetilde{\rho}\right\}  \right]  +\lambda_{\ell^{\prime}\ell}\left[
x_{\ell},\left\{  x_{\ell^{\prime}},\widetilde{\rho}\right\}  \right]
\right)  \right] \nonumber
\end{align}
where the Hamiltonian $\mathbf{H}_{i}$, is given by
\begin{equation}
\mathbf{H}_{i}=\frac{1}{2}\sum_{\ell}\left(  \frac{p_{\ell}^{2}}{m_{\ell}%
}+m_{\ell}\widetilde{\omega}_{\ell}^{2}x_{\ell}^{2}\right)  +\lambda_{11}%
x_{1}x_{2}+\lambda_{22}p_{1}p_{2}+\lambda_{21}p_{1}x_{2}+\lambda_{12}%
x_{1}p_{2}\text{.} \label{41}%
\end{equation}

Note that, when turning off the coupling between the two oscillators, we
obtain from Eq. (\ref{40.5}) two independent dissipative oscillators described
by two independent CL models. Therefore, in the case of distinct reservoirs,
there is no effective coupling induced between the oscillators.

\subsection{\textbf{A Common\ Reservoir}}

For the case of a common reservoir, we obtain the influence functional%
\begin{align}%
\mathcal{F}%
\left[  \left\{  x_{\ell}\right\}  ,\left\{  y_{\ell}\right\}  \right]   &
=\int\operatorname{d}\mathbf{X}\operatorname{d}\mathbf{X}^{\prime
}\operatorname{d}\mathbf{Y}^{\prime}\rho_{R}\left(  \mathbf{X}^{\prime
},\mathbf{Y}^{\prime},0\right)
{\displaystyle\int}
\operatorname{D}\mathbf{X(\tau)}\operatorname{D}\mathbf{Y(\tau)}\nonumber\\
&  {\times}\exp\left\{  \frac{i}{\hbar}\left[  \mathcal{S}_{R}\left[
\mathbf{X}\right]  -\mathcal{S}_{R}\left[  \mathbf{Y}\right]  +\sum_{\ell
}\left(  \mathcal{S}_{S_{\ell}/R}\left[  x_{\ell},\mathbf{X}\right]
-\mathcal{S}_{S_{\ell}/R}\left[  y_{\ell},\mathbf{Y}\right]  \right)  \right]
\right\}  \text{,} \label{42}%
\end{align}
which differs from that of the CL model for an additional system-reservoir
coupling. Assuming identical system-reservoir couplings $C_{1k}=C_{2k}=C_{k}$,
the solution of the influence functional (\ref{24}), following directly from
that in Refs. \cite{22,23}, is given by%

\begin{align}%
\mathcal{F}%
\left[  \left\{  x_{\ell}\right\}  ,\left\{  y_{\ell}\right\}  \right]   &
=\exp\left(  -\frac{1}{\hbar}\sum_{\ell}\int_{0}^{t}\operatorname{d}\tau
\int_{0}^{\tau}\operatorname{d}t^{\prime}\text{ }\left[  x_{\ell}%
(\tau)-y_{\ell}(\tau)\right]  \alpha_{\operatorname{R}}(\tau-t^{\prime
})\left[  x_{\ell}(t^{\prime})-y_{\ell}(t^{\prime})\right]  \right)
\nonumber\\
&  \times\exp\left(  -\frac{i}{\hbar}\sum_{\ell}\int_{0}^{t}\operatorname{d}%
\tau\int_{0}^{\tau}\operatorname{d}t^{\prime}\text{ }\left[  x_{\ell}%
(\tau)-y_{\ell}(\tau)\right]  \alpha_{\operatorname{I}}(\tau-t^{\prime
})\left[  x_{\ell}(t^{\prime})+y_{\ell}(t^{\prime})\right]  \right)  \text{,}
\label{43}%
\end{align}
where the real and imaginary parts of a function $\alpha\left(  t\right)  $
read
\begin{subequations}
\label{44}%
\begin{align}
\alpha_{\operatorname{R}}(t)  &  =\sum_{k}\frac{C_{k}^{2}}{2m_{k}\omega_{k}%
}\coth\left(  \omega\hbar\beta/2\right)  \cos\left[  \omega(t)\right]
\text{,}\label{44a}\\
\alpha_{\operatorname{I}}(t)  &  =-\sum_{k}\frac{C_{k}^{2}}{2m_{k}\omega_{k}%
}\sin\left[  \omega(t)\right]  \text{.} \label{44b}%
\end{align}

Defining, as in the case of distinct reservoirs, a spectral function%
\end{subequations}
\begin{equation}
\chi(\omega)=\chi_{\ell}(\omega)=\sum_{k}\frac{\pi C_{k}^{2}}{2m_{k}\omega
_{k}}\delta(\omega-\omega_{k}) \label{45}%
\end{equation}
constrained to the range defined by a frequency cutoff $\Omega^{C}$, and also
a damping constant $\eta$, we obtain for an Ohmic reservoir
\begin{equation}
\chi(\omega)=\left\{
\begin{array}
[c]{cc}%
\eta\omega\text{,} & \omega<\Omega^{C}\text{,}\\
0\text{,} & \omega>\Omega^{C}\text{.}%
\end{array}
\right.  \label{46}%
\end{equation}
\qquad

Using exactly the same approximations performed for the case of distinct
reservoirs, we obtain the influence functional%
\begin{align}
&
\mathcal{F}%
\left[  \left\{  x_{\ell}\right\}  ,\left\{  y_{\ell}\right\}  \right]
\nonumber\\
&  =\exp\left\{  -\frac{i}{\hbar}\left[  F\left(  \left\{  x_{\ell}\right\}
,\left\{  x_{\ell}^{\prime}\right\}  ,\left\{  y_{\ell}\right\}  ,\left\{
y_{\ell}^{\prime}\right\}  \right)  -\frac{\eta\Omega^{C}}{\pi}\sum_{\ell
,\ell^{\prime}}\int_{0}^{t}\operatorname{d}\tau\text{ }\left(  x_{\ell}%
(\tau)x_{\ell^{\prime}}(\tau)-y_{\ell}(\tau)y_{\ell^{\prime}}(\tau)\right)
\right.  \right. \nonumber\\
&  \left.  \left.  +\mu\gamma\int_{0}^{t}\operatorname{d}\tau\text{ }%
\sum_{\ell}\left(  \sum_{\ell^{\prime}}\left(  x_{\ell}(\tau)\dot{y}%
_{\ell^{\prime}}(\tau)-y_{\ell}(\tau)\dot{x}_{\ell^{\prime}}(\tau)\right)
+\sum_{\ell^{\prime}\left(  \neq\ell\right)  }\left(  x_{\ell}(\tau)\dot
{x}_{\ell^{\prime}}(\tau)-y_{\ell}(\tau)\dot{y}_{\ell^{\prime}}(\tau)\right)
\right)  \right]  \right\} \nonumber\\
&  \times\exp\left[  -\frac{2\mu\gamma}{\pi\hbar}\sum_{\ell,\ell^{\prime}}%
\int_{0}^{t}\operatorname{d}\tau\int_{0}^{\tau}\operatorname{d}t^{\prime
}\text{ }\left[  x_{\ell}(\tau)-y_{\ell}(\tau)\right]  \left[  x_{\ell
^{\prime}}(t^{\prime})-y_{\ell^{\prime}}(t^{\prime})\right]  \right.
\nonumber\\
&  \left.  \times\int_{0}^{\Omega^{C}}\operatorname{d}\omega\text{ }%
\omega\coth\left(  \omega\hbar\beta/2\right)  \cos\left[  \omega
(\tau-t^{\prime})\right]  \right]  \label{47}%
\end{align}
where
\begin{equation}
F\left(  \left\{  x_{\ell}\right\}  ,\left\{  x_{\ell}^{\prime}\right\}
,\left\{  y_{\ell}\right\}  ,\left\{  y_{\ell}^{\prime}\right\}  \right)
=\frac{\mu\gamma}{2}\sum_{\ell}\left[  \left(  x_{\ell}^{2}-x_{\ell}^{\prime
2}\right)  -\left(  y_{\ell}^{2}-y_{\ell}^{\prime2}\right)  \right]  \text{.}
\label{47l}%
\end{equation}
We stress that we have defined, after the assumption $\left\{  C_{\ell
k}\right\}  =C_{k}$, the\ relaxation constant $\gamma=\eta/2\mu$ for both
system-reservoir couplings, which implies immediately that $\left\{  \mu
_{\ell\ell}\right\}  =\mu$ and, consequently, $\left\{  m_{\ell}\right\}  =m$.
For the master equation, we obtain
\begin{align}
\frac{\partial\widetilde{\rho}}{\partial t}  &  =-\sum_{\ell}\left\{
-i\frac{\hbar}{2m}\left(  \frac{\partial^{2}}{\partial x_{\ell}^{2}}%
-\frac{\partial^{2}}{\partial y_{\ell}^{2}}\right)  +i\frac{m\widetilde
{\omega}_{\ell}^{2}}{2\hbar}\left(  x_{\ell}^{2}-y_{\ell}^{2}\right)  \right.
\nonumber\\
&  +\sum_{\ell^{\prime}}\left(  x_{\ell^{\prime}}-y_{\ell^{\prime}}\right)
\left[  \frac{2m\widetilde{\gamma}k_{B}T}{\hbar^{2}}\left(  x_{\ell}-y_{\ell
}\right)  +\widetilde{\gamma}\left(  \frac{\partial}{\partial x_{\ell}}%
-\frac{\partial}{\partial y_{\ell}}\right)  \right] \nonumber\\
&  +\sum_{\ell^{\prime}\left(  \neq\ell\right)  }\left[  -i\frac{\hbar
\lambda_{22}}{2}\left(  \frac{\partial^{2}}{\partial x_{\ell}\partial
x_{\ell^{\prime}}}-\frac{\partial^{2}}{\partial y_{\ell}\partial
y_{\ell^{\prime}}}\right)  +\lambda_{\ell^{\prime}\ell}\left(  x_{\ell
^{\prime}}\frac{\partial}{\partial x_{\ell}}+y_{\ell^{\prime}}\frac{\partial
}{\partial y_{\ell}}\right)  \right. \nonumber\\
&  \left.  \left.  +\frac{i}{2\hbar}\widetilde{\lambda}_{11}\left(  x_{\ell
}x_{\ell^{\prime}}-y_{\ell}y_{\ell^{\prime}}\right)  +i\frac{m\widetilde
{\gamma}\lambda_{\ell\ell^{\prime}}}{\hbar}\left(  x_{\ell}+y_{\ell}\right)
\sum_{\ell^{\prime\prime}}\left(  x_{\ell^{\prime\prime}}-y_{\ell
^{\prime\prime}}\right)  \right]  \right\}  \widetilde{\rho}\text{,}
\label{48}%
\end{align}
in which the renormalized damping constant $\widetilde{\gamma}=\gamma/\left(
1-m^{2}\lambda_{22}^{2}\right)  $ and the effective coupling
parameter$\widetilde{\lambda}_{11}=\lambda_{11}-2\Omega^{C}\eta/\pi\ $are
considered. As usual, this shift for the effective coupling parameter is small
and can be included in $\lambda_{11}.$ The operator equation associated with
the above coordinate representation turns out to be%
\begin{align}
\frac{\partial\widetilde{\rho}}{\partial t}  &  =-\frac{i}{\hbar}\left[
\mathbf{H}_{ii},\widetilde{\rho}\right]  -i\frac{m\widetilde{\gamma}}{\hbar
}\sum_{\ell,\ell^{\prime}\left(  \neq\ell\right)  }\lambda_{\ell\ell^{\prime}%
}\left[  x_{\ell^{\prime}},\left\{  x_{\ell},\widetilde{\rho}\right\}  \right]
\nonumber\\
&  -\sum_{\ell,\ell^{\prime}}\left[  i\frac{\widetilde{\gamma}}{\hbar}\left[
x_{\ell},\left\{  p_{\ell^{\prime}},\widetilde{\rho}\right\}  \right]
+\frac{2m\widetilde{\gamma}k_{B}T}{\hbar^{2}}\left[  x_{\ell},\left[
x_{\ell^{\prime}},\widetilde{\rho}\right]  \right]  \right]  \label{49}%
\end{align}
where the Hamiltonian $\mathbf{H}_{ii}$, given by
\begin{equation}
\mathbf{H}_{ii}=\frac{1}{2}\sum_{\ell}\left(  \frac{p_{\ell}^{2}}{m}%
+m\varpi_{\ell}^{2}x_{\ell}^{2}\right)  +\lambda_{11}x_{1}x_{2}+\lambda
_{22}p_{1}p_{2}+\lambda_{12}x_{1}p_{2}+\lambda_{21}p_{1}x_{2}\text{,}
\label{50}%
\end{equation}
contains the renormalized frequencies
\begin{equation}
\varpi_{\ell}^{2}=\widetilde{\omega}_{\ell}^{2}+2\widetilde{\gamma}\left(
\lambda_{12}\delta_{\ell1}+\lambda_{21}\delta_{\ell2}\right)  \text{.}
\label{51}%
\end{equation}

\textbf{A reservoir-induced coupling between the oscillators}

It is worth stressing that, differently from the case of distinct reservoirs
presented in Eq. (\ref{40}), the common reservoir induces an effective
coupling between the two oscillators, even when their original interactions
$\left\{  \lambda_{\ell\ell^{\prime}}\right\}  $ are turned off. In fact, with
$\lambda_{\ell\ell^{\prime}}=0$, the Eq. (\ref{49}) simplifies to
\begin{equation}
\frac{\partial\widetilde{\rho}}{\partial t}=-\frac{i}{\hbar}%
{\textstyle\sum\limits_{\ell}}
\left[  H_{S_{\ell}},\widetilde{\rho}\right]  -\sum_{\ell,\ell^{\prime}%
}\left[  i\frac{\gamma}{\hbar}\left[  x_{\ell},\left\{  p_{\ell^{\prime}%
},\widetilde{\rho}\right\}  \right]  +\frac{2m\gamma k_{B}T}{\hbar^{2}}\left[
x_{\ell},\left[  x_{\ell^{\prime}},\widetilde{\rho}\right]  \right]  \right]
\text{,}%
\end{equation}
giving two independent CL models, for $\ell=\ell^{\prime}$, apart from the
reservoir-induced coupling between the oscillators, for $\ell\neq\ell^{\prime
}$. This effective coupling, for the case analyzed here of an ohmic reservoir
in the high-temperature regime, consists therefore of a dissipative and a
diffusive term, given by $\left(  \gamma/i\hbar\right)  \left(  \left[
x_{1},\left\{  p_{2},\widetilde{\rho}\right\}  \right]  +\left[
x_{2},\left\{  p_{1},\widetilde{\rho}\right\}  \right]  \right)  $ and
$\left(  2m\gamma k_{B}T/\hbar^{2}\right)  \left[  x_{\ell},\left[
x_{\ell^{\prime}},\widetilde{\rho}\right]  \right]  $, respectively.
Interestingly, such dissipative and diffusive terms couple together the
variables of both oscillators while, evidently, the equivalent terms in the CL
model apply to the variables of a single particle.

\section{Solutions of the Master Equations (\ref{39}) and (\ref{48})}

In this section we present the solutions of the master equations governing the
dynamics of the coupled dissipative harmonic oscillators in both cases of
separate reservoirs and a common one. These solutions enable us to analyze the
coherence and decoherence dynamics of quantum superpositions prepared in one
of the oscillators of entangled states prepared in both oscillators. Moreover,
the form of the solutions presented here enable a complete understanding of
the evolution of such states, thus enlarging the perspective of the coherence
and decoherence analysis offered by the CL model \cite{PRA}.

\subsection{\textbf{Distinct\ Reservoirs}}

The introduction of the collective\ ($R_{\ell}$) and relative ($r_{\ell}$)
coordinates
\begin{equation}
R_{\ell}=\frac{x_{\ell}+y_{\ell}}{2}\qquad\text{,}\qquad r_{\ell}=x_{\ell
}-y_{\ell} \label{52}%
\end{equation}
enable us to rewrite the master equation (\ref{39}) in the form%
\begin{align}
\frac{\partial\widetilde{\rho}}{\partial t}  &  =\sum_{\ell}\left\{  \left(
-1\right)  ^{\ell}i\frac{\hbar}{m_{\ell}}\frac{\partial^{2}}{\partial r_{\ell
}\partial R_{\ell}}-2\widetilde{\gamma}_{\ell}r_{\ell}\frac{\partial}{\partial
r_{\ell}}+i\frac{m_{\ell}\omega_{\ell}^{2}}{\hbar}R_{\ell}r_{\ell}%
-\frac{D_{\ell}}{\hbar^{2}}r_{\ell}^{2}\right. \nonumber\\
&  \left.  -\sum_{\ell^{\prime}\left(  \neq\ell\right)  }\left[  \left(
\lambda_{\ell\ell^{\prime}}R_{\ell}\frac{\partial}{\partial R_{\ell^{\prime}}%
}+\Delta_{\ell^{\prime}}r_{\ell^{\prime}}\frac{\partial}{\partial r_{\ell}%
}\right)  +i\hbar\lambda_{22}\frac{\partial^{2}}{\partial r_{\ell}\partial
R_{\ell^{\prime}}}-i\frac{\Gamma_{\ell}}{\hbar}R_{\ell^{\prime}}r_{\ell
}\right]  \right\}  \widetilde{\rho}\text{,} \label{53}%
\end{align}
where the diffusion coefficients $D_{\ell}$ and the effective coupling
parameters $\Gamma_{\ell}$ and $\Delta_{\ell}$ are given by
\begin{subequations}
\begin{align}
D_{\ell}  &  =2m_{\ell}\widetilde{\gamma}_{\ell}k_{B}T_{\ell}\text{,}%
\label{54a}\\
\Gamma_{\ell}  &  =\lambda_{11}+2m_{\ell}\widetilde{\gamma}_{\ell}\sum
_{\ell^{\prime}\left(  \neq\ell\right)  }\lambda_{\ell^{\prime}\ell}%
\text{,}\label{54b}\\
\Delta_{\ell}  &  =2\lambda_{22}m_{\ell}\widetilde{\gamma}_{\ell}+\sum
_{\ell^{\prime}\left(  \neq\ell\right)  }\lambda_{\ell\ell^{\prime}}\text{.}
\label{54c}%
\end{align}

By the partial Fourier transform%

\end{subequations}
\begin{equation}
\widetilde{\rho}\left(  \left\{  K_{\ell}\right\}  ,\left\{  r_{\ell}\right\}
,t\right)  =\frac{1}{2\pi}\left(  \prod_{\ell}\int_{-\infty}^{+\infty
}\operatorname{d}R_{\ell}\operatorname{e}^{-iK_{\ell}R_{\ell}}\right)
\widetilde{\rho}\left(  \left\{  R_{\ell}\right\}  ,\left\{  r_{\ell}\right\}
,t\right)  \label{55}%
\end{equation}
we reduce the second order partial differential equation (\ref{53}) to the
first order one%

\begin{align}
\frac{\partial\widetilde{\rho}}{\partial t}  &  =-\sum_{\ell}\left\{  \left(
2\widetilde{\gamma}_{\ell}r_{\ell}-\frac{\hbar}{m_{\ell}}K_{\ell}\right)
\frac{\partial}{\partial r_{\ell}}+\frac{m_{\ell}\omega_{\ell}^{2}}{\hbar
}r_{\ell}\frac{\partial}{\partial K_{\ell}}+\frac{D_{\ell}}{\hbar^{2}}r_{\ell
}^{2}\right. \nonumber\\
&  \left.  +\sum_{\ell^{\prime}\left(  \neq\ell\right)  }\left[  \left(
\frac{\Gamma_{\ell^{\prime}}r_{\ell^{\prime}}}{\hbar}-\lambda_{\ell
\ell^{\prime}}K_{\ell^{\prime}}\right)  \frac{\partial}{\partial K_{\ell}%
}+\left(  \Delta_{\ell^{\prime}}r_{\ell^{\prime}}-\hbar\lambda_{22}%
K_{\ell^{\prime}}\right)  \frac{\partial}{\partial r_{\ell}}\right]  \right\}
\widetilde{\rho} \label{56}%
\end{align}
whose solution can be obtained by the method of characteristics
\cite{Courant,Venugopalan}. Defining the curves%

\begin{equation}
K_{\ell}=K_{\ell}(s)\text{, }r_{\ell}=r_{\ell}(s)\text{, and }t=t(s)
\label{57}%
\end{equation}
we obtain, from the partial differential equation (\ref{56}), a system of
coupled ordinary differential equations%

\begin{subequations}
\begin{align}
\frac{\operatorname*{d}r_{\ell}}{\operatorname*{d}s}  &  =-\hbar\sum
_{\ell^{\prime}}\left[  \lambda_{22}\left(  1-\delta_{\ell\ell^{\prime}%
}\right)  +\frac{1}{m_{\ell}}\delta_{\ell\ell^{\prime}}\right]  K_{\ell
^{\prime}}+2\widetilde{\gamma}_{\ell}r_{\ell}+\left(  \Delta_{1}r_{1}%
\delta_{\ell2}+\Delta_{2}r_{2}\delta_{\ell1}\right)  \text{,}\label{58a}\\
\frac{\operatorname*{d}K_{\ell}}{\operatorname*{d}s}  &  =\frac{m_{\ell}%
\omega_{\ell}^{2}}{\hbar}r_{\ell}+\frac{1}{\hbar}\left[  \left(  \Gamma
_{2}r_{2}-\hbar\lambda_{12}K_{2}\right)  \delta_{\ell1}+\left(  \Gamma
_{1}r_{1}-\hbar\lambda_{21}K_{1}\right)  \delta_{\ell2}\right]  \text{,}%
\label{58b}\\
\frac{\operatorname*{d}\widetilde{\rho}}{\operatorname*{d}s}  &  =-\frac
{1}{\hbar^{2}}\sum_{\ell}D_{\ell}r_{\ell}^{2}\widetilde{\rho}\text{,}%
\label{58c}\\
\frac{\operatorname*{d}t}{\operatorname*{d}s}  &  =1\text{.} \label{58d}%
\end{align}

The first four equations of the above system can be expressed, in matrix form, as%

\end{subequations}
\begin{equation}
\frac{\operatorname*{d}}{\operatorname*{d}t}\left(
\begin{array}
[c]{c}%
r_{1}\left(  t\right) \\
K_{1}\left(  t\right) \\
r_{2}\left(  t\right) \\
K_{2}\left(  t\right)
\end{array}
\right)  =\left(
\begin{array}
[c]{cccc}%
2\widetilde{\gamma}_{1} & -\frac{\hbar}{m_{1}} & \Delta_{2} & -\hbar
\lambda_{22}\\
\frac{1}{\hbar}m_{1}\omega_{1}^{2} & 0 & \frac{1}{\hbar}\Gamma_{2} &
-\lambda_{12}\\
\Delta_{1} & -\hbar\lambda_{22} & 2\widetilde{\gamma}_{2} & -\frac{\hbar
}{m_{2}}\\
\frac{\Gamma_{1}}{\hbar} & -\lambda_{21} & \frac{1}{\hbar}m_{2}\omega_{2}^{2}
& 0
\end{array}
\right)  \left(
\begin{array}
[c]{c}%
r_{1}\left(  t\right) \\
K_{1}\left(  t\right) \\
r_{2}\left(  t\right) \\
K_{2}\left(  t\right)
\end{array}
\right)  \text{,} \label{59}%
\end{equation}
where we shall denote the square matrix by $\mathbf{M}$ and its eigenvalues by
$\Lambda_{m}$. Next, we note that the solution of this system of coupled
equations can be written in the form%

\begin{equation}
\left(
\begin{array}
[c]{c}%
r_{1}\left(  t\right) \\
K_{1}\left(  t\right) \\
r_{2}\left(  t\right) \\
K_{2}\left(  t\right)
\end{array}
\right)  =\left(
\begin{array}
[c]{cccc}%
\eta_{11} & \eta_{12} & \eta_{13} & \eta_{14}\\
\eta_{21} & \eta_{22} & \eta_{23} & \eta_{24}\\
\eta_{31} & \eta_{32} & \eta_{33} & \eta_{34}\\
\eta_{41} & \eta_{42} & \eta_{43} & \eta_{44}%
\end{array}
\right)  \left(
\begin{array}
[c]{c}%
c_{1}\left(  t\right) \\
c_{2}\left(  t\right) \\
c_{3}\left(  t\right) \\
c_{4}\left(  t\right)
\end{array}
\right)  \text{,} \label{60}%
\end{equation}
where the eigenvector $\left(
\begin{array}
[c]{cccc}%
\eta_{1n} & \eta_{2n} & \eta_{3n} & \eta_{4n}%
\end{array}
\right)  ^{\top}$ is associated with the eigenvalue $\Lambda_{n}$.

Making use of both solutions for $r_{\ell}\left(  t\right)  $ and that for the
third ordinary differential equation (\ref{58c}), given by%

\begin{equation}
\widetilde{\rho}\left(  \left\{  K_{\ell}\right\}  ,\left\{  r_{\ell}\right\}
,t\right)  =B\exp\left[  -\frac{1}{\hbar^{2}}\sum_{\ell}D_{\ell}%
\int\operatorname*{d}t\ r_{\ell}^{2}\left(  t\right)  \right]  \text{,}
\label{64}%
\end{equation}
we finally obtain the density matrix%

\begin{equation}
\widetilde{\rho}\left(  \left\{  K_{\ell}\right\}  ,\left\{  r_{\ell}\right\}
,t\right)  =\widetilde{\rho}\left(  \left\{  K_{\ell}^{\prime}\right\}
,\left\{  r_{\ell}^{\prime}\right\}  ,0\right)  \operatorname*{e}%
\nolimits^{-Z\left(  \left\{  K_{\ell}\right\}  ,\left\{  r_{\ell}\right\}
,t\right)  /\hbar^{2}}\text{,} \label{65}%
\end{equation}
with%
\begin{equation}
Z\left(  \left\{  K_{\ell}\right\}  ,\left\{  r_{\ell}\right\}  ,t\right)
=\sum_{m,n}c_{m}\left(  t\right)  c_{n}\left(  t\right)  \frac{D_{1}\eta
_{1m}\eta_{1n}+D_{2}\eta_{3m}\eta_{3n}}{\Lambda_{m}+\Lambda_{n}}\left(
1-\operatorname{e}^{-\left(  \Lambda_{m}+\Lambda_{n}\right)  t}\right)
\text{,} \label{66}%
\end{equation}
and $c_{m}\left(  t\right)  =c_{m}\left(  0\right)  e^{\Lambda_{m}t}$, where
$c_{m}\left(  0\right)  $ is determined by the initial conditions.

The next step is to calculate the inverse Fourier transform of Eq. (\ref{55}).
To this end, let us suppose that the coupled harmonic oscillators are prepared
at $t=0$ in the general superposition of Gaussian functions of width $\left\{
\sigma_{\ell}\right\}  $:%

\begin{equation}
\Psi\left(  \left\{  x_{\ell}\right\}  ,0\right)  =\sum_{\ell}P_{\ell}%
\prod_{\ell^{\prime}}\operatorname*{e}\nolimits^{-\left[  \left(
x_{\ell^{\prime}}+q_{\ell^{\prime}\ell}\right)  /\sigma_{\ell^{\prime}%
}\right]  ^{2}}\text{,} \label{68}%
\end{equation}
which include both separable and entangled states, depending on the choice of
the Gaussian centers $q_{1\ell}$ and $q_{2\ell}$ for oscillators $1$ and $2$.
Rewriting the density matrix for this wave function, given by%

\begin{equation}
\widetilde{\rho}\left(  \left\{  x_{\ell}\right\}  ,\left\{  y_{\ell}\right\}
,0\right)  =\sum_{\ell,\ell^{\prime}}P_{\ell}P_{\ell^{\prime}}\prod
_{\ell^{\prime\prime}}\operatorname*{e}\nolimits^{-\left[  \left(
x_{\ell^{\prime\prime}}+q_{\ell^{\prime\prime}\ell}\right)  /\sigma
_{\ell^{\prime\prime}}\right]  ^{2}}\operatorname*{e}\nolimits^{-\left[
\left(  y_{\ell^{\prime\prime}}+q_{\ell^{\prime\prime}\ell}\right)
/\sigma_{\ell^{\prime\prime}}\right]  ^{2}}\text{,} \label{69}%
\end{equation}
with the collective and relative coordinates defined in Eq. (\ref{52}), and
computing its Fourier transform, we obtain%

\begin{align}
&  \widetilde{\rho}\left(  \left\{  K_{\ell}^{\prime}\right\}  ,\left\{
r_{\ell}^{\prime}\right\}  ,0\right) \nonumber\\
&  =\frac{1}{4}%
{\displaystyle\sum\limits_{\ell^{\prime},\ell^{\prime\prime}}}
P_{\ell^{\prime}}P_{\ell^{\prime\prime}}\prod_{\ell}\sigma_{\ell}\exp\left\{
-2\left[  \left(  \frac{\sigma_{\ell}}{4}K_{\ell}^{\prime}\right)
^{2}-i\left(  q_{\ell\ell^{\prime}}+q_{\ell\ell^{\prime\prime}}\right)
K_{\ell}^{\prime}+\left(  \frac{q_{\ell\ell^{\prime}}-q_{\ell\ell
^{\prime\prime}}+r_{\ell}^{\prime}}{2\sigma_{\ell}}\right)  ^{2}\right]
\right\}  \text{.} \label{70}%
\end{align}

Therefore, from the results in Eqs. (\ref{65}) and (\ref{70}), we obtain the
general solution for the transformed master equation (\ref{56})%
\begin{align}
\widetilde{\rho}\left(  \left\{  K_{\ell}\right\}  ,\left\{  r_{\ell}\right\}
,t\right)   &  =\mathcal{R}\left(  \left\{  K_{\ell}\right\}  ,\left\{
r_{\ell}\right\}  ,t\right)
{\displaystyle\sum\limits_{\ell,\ell^{\prime}}}
\Upsilon_{\ell\ell^{\prime}}\nonumber\\
&  \times\exp\left\{  -\left[  \theta_{\ell\ell^{\prime}}^{\left(  1\right)
}(t)r_{1}+\theta_{\ell\ell^{\prime}}^{\left(  2\right)  }(t)K_{1}+\theta
_{\ell\ell^{\prime}}^{\left(  3\right)  }(t)r_{2}+\theta_{\ell\ell^{\prime}%
}^{\left(  4\right)  }(t)K_{2}\right]  \right\} \nonumber\\
&  \times\exp\left\{  i\left[  \widetilde{\theta}_{\ell\ell^{\prime}}^{\left(
1\right)  }(t)r_{1}+\widetilde{\theta}_{\ell\ell^{\prime}}^{\left(  2\right)
}(t)K_{1}+\widetilde{\theta}_{\ell\ell^{\prime}}^{\left(  3\right)  }%
(t)r_{2}+\widetilde{\theta}_{\ell\ell^{\prime}}^{\left(  4\right)  }%
(t)K_{2}\right]  \right\}  \label{72}%
\end{align}
which, for $t=0$, gives the initial condition (\ref{69}). Assuming that the
matrix elements $\varepsilon_{mn}$ follow from the inverse of the matrix
composed by $\eta_{mn}$, we have defined the functions%
\begin{align}
\mathcal{R}\left(  \left\{  K_{\ell}\right\}  ,\left\{  r_{\ell}\right\}
,t\right)   &  =\exp\left\{  -\left[  \Phi_{11}(t)r_{1}^{2}+\Phi_{33}%
(t)r_{2}^{2}+\Phi_{13}(t)r_{1}r_{2}+\Phi_{12}(t)r_{1}K_{1}+\Phi_{14}%
(t)r_{1}K_{2}\right.  \right. \nonumber\\
&  \left.  \left.  +\Phi_{22}(t)K_{1}^{2}+\Phi_{44}(t)K_{2}^{2}+\Phi
_{24}(t)K_{1}K_{2}+\Phi_{23}(t)r_{2}K_{1}+\Phi_{34}(t)r_{2}K_{2}\right]
\right\}
\end{align}

\begin{subequations}
\label{73}%
\begin{align}
\Upsilon_{\ell\ell^{\prime}}  &  =\frac{1}{4}P_{\ell}P_{\ell^{\prime}}%
\prod_{\ell^{\prime\prime}}\sigma_{\ell^{\prime\prime}}\exp\left[  -\frac
{1}{2}\left(  \frac{q_{\ell^{\prime\prime}\ell}-q_{\ell^{\prime\prime}%
\ell^{\prime}}}{\sigma_{\ell^{\prime\prime}}}\right)  ^{2}\right]
\text{,}\label{73a}\\
\Phi_{kk^{\prime}}(t)  &  =\sum_{i,j=1(j\geq i)}^{4}\left(  -1\right)
^{i+j}2^{-\delta_{ij}}\left[  \left(  \frac{\eta_{1i}\eta_{1j}}{\sigma_{1}%
^{2}}+\frac{\sigma_{1}^{2}}{4}\eta_{2i}\eta_{2j}+\frac{\eta_{3i}\eta_{3j}%
}{\sigma_{2}^{2}}+\frac{\sigma_{2}^{2}}{4}\eta_{4i}\eta_{4j}\right)
\operatorname{e}^{-\left(  \Lambda_{i}+\Lambda_{j}\right)  t}\right.
\nonumber\\
&  \left.  +2\zeta_{ij}\left(  1-\operatorname{e}^{-(\Lambda_{1}+\Lambda
_{2})t}\right)  \right]  \left[  \varepsilon_{ik}\varepsilon_{jk^{\prime}%
}-\left(  1-\delta_{kk^{\prime}}\right)  \varepsilon_{ik^{\prime}}%
\varepsilon_{jk}\right]  \text{,}\label{73b}\\
\theta_{\ell\ell^{\prime}}^{\left(  k\right)  }(t)  &  =\left(  -1\right)
^{\ell+1}\sum_{j=1}^{4}\left(  \frac{q_{1\ell}-q_{1\ell^{\prime}}}{\sigma
_{1}^{2}}\eta_{1j}+\frac{q_{2\ell}-q_{2\ell^{\prime}}}{\sigma_{2}^{2}}%
\eta_{3j}\right)  \varepsilon_{jk}\operatorname{e}^{-\Lambda_{j}t}%
\text{,}\label{73c}\\
\widetilde{\theta}_{\ell\ell^{\prime}}^{\left(  k\right)  }(t)  &  =\left(
-1\right)  ^{\ell+1}\sum_{j=1}^{4}\left(  \frac{q_{1\ell}+q_{1\ell^{\prime}}%
}{2}\eta_{2j}+\frac{q_{2\ell}+q_{2\ell^{\prime}}}{2}\eta_{4j}\right)
\varepsilon_{jk}\operatorname{e}^{-\Lambda_{j}t}. \label{73d}%
\end{align}

The next and final step is to calculate the inverse Fourier transform of Eq.
(\ref{72}). At this point we observe that it is highly recommendable to write
the Fourier transformed density matrix $\widetilde{\rho}\left(  \left\{
R_{\ell}\right\}  ,\left\{  r_{\ell}\right\}  ,0\right)  $ in its initial
normalized form, obtained by a coordinate transformation of Eq. (\ref{69}):%
\end{subequations}
\begin{align}
\widetilde{\rho}\left(  \left\{  R_{\ell}\right\}  ,\left\{  r_{\ell}\right\}
,0\right)   &  =\left\{  \pi\sigma_{1}\sigma_{2}\left[  1+\prod_{\ell}%
\exp\left[  -\frac{1}{2}\left(  \frac{q_{\ell1}-q_{\ell2}}{\sigma_{\ell}%
}\right)  ^{2}\right]  \right]  \right\}  ^{-1}\nonumber\\
&  \times\sum_{\ell^{\prime},\ell^{\prime\prime}}\exp\left[  -\sum_{\ell}%
\frac{\left(  q_{\ell\ell^{\prime}}-q_{\ell\ell^{\prime\prime}}+r_{\ell
}\right)  ^{2}+\left(  q_{\ell\ell^{\prime}}+q_{\ell\ell^{\prime\prime}%
}+2R_{\ell}\right)  ^{2}}{2\sigma_{\ell}^{2}}\right]  \text{.} \label{74}%
\end{align}

The exponentials in Eq. (\ref{74}) associated with $\ell=\ell^{\prime}$,
represent the diagonal elements of $\widetilde{\rho}\left(  \left\{  R_{\ell
}\right\}  ,\left\{  r_{\ell}\right\}  ,0\right)  $, which are associated with
the probability amplitudes of the system, while exponentials associated with
$\ell\neq\ell^{\prime}$ represent the off-diagonal elements, associated with
phase coherence. After the sudden system-reservoir couplings it is expected
that the diagonal elements will be dragged to the origin of the coordinates by
the dissipative mechanisms, whereas the off-diagonal elements will vanish
continuously, owing to the associated noise injection into the system.
Therefore, by keeping exactly the form of the above initial density matrix
(\ref{74}) after its time evolution, we directly verify such expected
dynamics, simplifying our evaluation of the decoherence effects. From this
perspective, we obtain the final solution%

\begin{align}
\widetilde{\rho}\left(  \left\{  R_{\ell}\right\}  ,\left\{  r_{\ell}\right\}
,t\right)   &  =\left\{  4\pi\Sigma(t)\left[  1+\prod_{\ell}\exp\left[
-\frac{1}{2}\left(  \frac{q_{\ell1}-q_{\ell2}}{\sigma_{\ell}}\right)
^{2}\right]  \right]  \right\}  ^{-1}\nonumber\\
&  \times\sum_{\ell,\ell^{\prime}}\exp\left\{  -\frac{1}{2}\sum_{\ell}\left(
1-\delta_{\ell\ell^{\prime}}\right)  \left(  \frac{q_{\ell1}-q_{\ell2}}%
{\sigma_{\ell}}\right)  ^{2}\left[  1-\Gamma(t)\right]  \right\} \nonumber\\
&  \times\exp\left[  -\frac{1}{2}\Xi_{\ell\ell^{\prime}}\left(  \left\{
R_{\ell}\right\}  ,\left\{  r_{\ell}\right\}  ,t\right)  +i\Theta_{\ell
\ell^{\prime}}\left(  \left\{  R_{\ell}\right\}  ,\left\{  r_{\ell}\right\}
,t\right)  \right]  \text{,} \label{75}%
\end{align}
where the functions $\Xi_{mn}\left(  \left\{  R_{\ell}\right\}  ,\left\{
r_{\ell}\right\}  ,t\right)  $, accounting for the dynamics of the diagonal
and off-diagonal elements, are given by
\begin{align}
\Xi_{mn}\left(  \left\{  R_{\ell}\right\}  ,\left\{  r_{\ell}\right\}
,t\right)   &  =\sum_{\ell,\ell^{\prime}}\left\{  2^{\delta_{\ell\ell^{\prime
}}+1}\left(  -1\right)  ^{\ell+\ell^{\prime}}\frac{R_{\ell}+\theta
_{mn}^{\left(  2\ell\right)  }(t)}{\Sigma_{\ell\ell^{\prime}}(t)}\frac
{R_{\ell^{\prime}}+\theta_{mn}^{\left(  2\ell^{\prime}\right)  }(t)}%
{\Sigma_{\ell^{\prime}\ell}(t)}\right. \nonumber\\
&  \left.  +2^{\delta_{\ell\ell^{\prime}}-1}\frac{r_{\ell}-\left[  2\left(
q_{\ell1}(t)-q_{\ell2}(t)\right)  \right]  ^{\delta_{\ell\ell^{\prime}}}%
}{\sigma_{\ell\ell^{\prime}}(t)}\frac{r_{\ell^{\prime}}-\left[  2\left(
q_{\ell^{\prime}1}(t)-q_{\ell^{\prime}2}(t)\right)  \right]  ^{\delta
_{\ell\ell^{\prime}}}}{\sigma_{\ell^{\prime}\ell}(t)}\right\}  \text{,}
\label{76}%
\end{align}
whereas the functions
\begin{align}
\Theta_{mn}\left(  \left\{  R_{\ell}\right\}  ,\left\{  r_{\ell}\right\}
,t\right)   &  =\sum_{\ell}\left\{  2\left[  R_{\ell}+\theta_{mn}^{\left(
2\ell\right)  }(t)\right]  \left[  \left(  -2\right)  ^{\delta_{\ell1}}%
\frac{\Phi_{12}(t)r_{1}+\Phi_{23}(t)r_{2}+\widetilde{\theta}_{mn}^{\left(
2\right)  }(t)}{\Sigma_{\ell1}^{2}(t)}\right.  \right. \nonumber\\
&  \left.  \left.  +\left(  -2\right)  ^{\delta_{\ell2}}\frac{\Phi
_{14}(t)r_{1}+\Phi_{34}(t)r_{2}+\widetilde{\theta}_{mn}^{\left(  4\right)
}(t)}{\Sigma_{\ell2}^{2}(t)}\right]  +\theta_{mn}^{\left(  2\ell-1\right)
}(t)r_{\ell}\right\}  \text{,} \label{77}%
\end{align}
account only for an oscillatory dynamics. Moreover, we also have, as part of
the normalization factor%

\begin{equation}
\Sigma(t)=\sqrt{4\Phi_{44}(t)\Phi_{22}(t)-\Phi_{24}^{2}(t)}\text{,} \label{78}%
\end{equation}
the functions associated with the widths of the diagonal and off-diagonal
Gaussian peaks:%

\begin{subequations}
\label{79}%
\begin{align}
\Sigma_{11}(t)  &  =\sqrt{\frac{2\Sigma^{2}(t)}{\Phi_{44}(t)}}\label{79a}\\
\Sigma_{22}(t)  &  =\sqrt{\frac{2\Sigma^{2}(t)}{\Phi_{22}(t)}}\label{79b}\\
\Sigma_{12}(t)  &  =\Sigma_{21}(t)=\sqrt{\frac{2\Sigma^{2}(t)}{\Phi_{24}(t)}%
}\label{79c}\\
\sigma_{11}(t)  &  =\sqrt{2\Phi_{11}(t)-2\frac{\Phi_{22}(t)\Phi_{14}%
^{2}(t)+\Phi_{44}(t)\Phi_{12}^{2}(t)-\Phi_{12}(t)\Phi_{14}(t)\Phi_{24}%
(t)}{\Sigma^{2}(t)}}\text{,}\label{79d}\\
\sigma_{22}(t)  &  =\sqrt{2\Phi_{33}(t)-2\frac{\Phi_{22}(t)\Phi_{34}%
^{2}(t)+\Phi_{44}(t)\Phi_{23}^{2}(t)-\Phi_{23}(t)\Phi_{34}(t)\Phi_{24}%
(t)}{\Sigma^{2}(t)}}\text{,}\label{79e}\\
\sigma_{12}(t)  &  =\sigma_{21}(t)=\left[  2\Phi_{13}(t)-2\frac{2\Phi
_{22}(t)\Phi_{14}(t)-\Phi_{12}(t)\Phi_{24}(t)}{\Sigma^{2}(t)}\Phi
_{34}(t)\right. \nonumber\\
&  \left.  -2\frac{2\Phi_{44}(t)\Phi_{12}(t)-\Phi_{14}(t)\Phi_{24}(t)}%
{\Sigma^{2}(t)}\Phi_{23}(t)\right]  ^{1/2}\text{.} \label{79f}%
\end{align}

Finally, regarding the exponential decay multiplying the off-diagonal elements
of the reduced density matrix (\ref{75}), we obtain the function%

\end{subequations}
\begin{equation}
\Gamma(t)=\left[  \sum_{\ell}\left(  \frac{q_{\ell1}-q_{\ell2}}{\sigma_{\ell}%
}\right)  ^{2}\right]  ^{-1}\sum_{\ell}\left[  \sigma_{\ell}^{2}%
(t)\vartheta_{\ell}^{2}(t)+\sum_{\ell^{\prime}}2^{\delta_{\ell\ell^{\prime}%
}+1}\left(  -1\right)  ^{\ell+\ell^{\prime}}\frac{\widetilde{\theta}%
_{12}^{\left(  2\ell\right)  }(t)\widetilde{\theta}_{12}^{\left(
2\ell^{\prime}\right)  }(t)}{\Sigma_{\ell\ell^{\prime}}^{2}(t)}\right]
\label{82}%
\end{equation}
where%

\begin{subequations}
\label{81}%
\begin{align}
\vartheta_{1}(t)  &  =\widetilde{\theta}_{1,2}^{\left(  1\right)  }%
(t)-\frac{\left[  2\Phi_{44}(t)\Phi_{12}(t)-\Phi_{14}(t)\Phi_{24}(t)\right]
\widetilde{\theta}_{12}^{\left(  2\right)  }(t)+\left[  2\Phi_{22}(t)\Phi
_{14}(t)-\Phi_{12}(t)\Phi_{24}(t)\right]  \widetilde{\theta}_{12}^{\left(
4\right)  }(t)}{\Sigma^{2}(t)}\text{,}\label{81a}\\
\vartheta_{2}(t)  &  =\widetilde{\theta}_{1,2}^{\left(  3\right)  }%
(t)-\frac{\left[  2\Phi_{44}(t)\Phi_{23}(t)-\Phi_{34}(t)\Phi_{24}(t)\right]
\widetilde{\theta}_{12}^{\left(  2\right)  }(t)+\left[  2\Phi_{22}(t)\Phi
_{34}(t)-\Phi_{23}(t)\Phi_{24}(t)\right]  \widetilde{\theta}_{12}^{\left(
4\right)  }(t)}{\Sigma^{2}(t)}\text{.} \label{81b}%
\end{align}

Comparing the initial and the evolved density matrices, given by Eqs.
(\ref{74}) and (\ref{75}), we verify, as expected, the displacement of both
the diagonal and off-diagonal peaks towards the origin of the coordinates,
described by Eq. (\ref{76}). Apart from this displacement towards the origin,
the system-reservoir coupling also induces the oscillatory term defined by Eq.
(\ref{77}).

\textbf{A general decay function of the off-diagonal peaks of the density
matrix}

From the evolved density matrix in Eq. (\ref{75}), we also deduce a general
expression for the decay $\mathcal{D}(t)$ of the off-diagonal peaks as time
goes on, represented by the exponential function
\end{subequations}
\begin{equation}
\mathcal{D}(t)=\exp\left\{  -2\left[  1-\Gamma(t)\right]  \sum_{\ell}\left[
\left(  q_{\ell1}-q_{\ell2}\right)  /2\sigma_{\ell}\right]  ^{2}\right\}  .
\label{83}%
\end{equation}
This decay function offers complete information relative to the decoherence of
any initial state of the coupled dissipative oscillators. We anticipate that
the same form of the decay function in Eq. (\ref{83}) will apply to the case
of a common reservoir, except for the time-dependent function $\Gamma(t)$,
which differs in the two cases as discussed below.

\subsection{\textbf{A Common\ Reservoir}}

The solution of the master equation for the case of a common reservoir is
entirely analogous to that of separate reservoirs. From the early assumption
of identical system-reservoir couplings $C_{1k}=C_{2k}=C_{k}$, rendering the
same masses $m$, damping rates $\gamma$, and diffusion coefficients $D$ for
both oscillators, the master equation, rewritten in terms of the
collective\ and relative coordinates in Eq. (\ref{52}), is given by%

\begin{align}
\frac{\partial\widetilde{\rho}}{\partial t}  &  =\sum_{\ell}\left[  \left(
-1\right)  ^{\ell}i\frac{\hbar}{m}\frac{\partial^{2}}{\partial r_{\ell
}\partial R_{\ell}}-2\widetilde{\gamma}r_{\ell}\frac{\partial}{\partial
r_{\ell}}+i\frac{m\omega_{\ell}^{2}}{\hbar}R_{\ell}r_{\ell}-\frac{D}{\hbar
^{2}}r_{\ell}^{2}\right. \nonumber\\
&  \left.  -\sum_{\ell^{\prime}\left(  \neq\ell\right)  }\left(  \lambda
_{\ell\ell^{\prime}}R_{\ell}\frac{\partial}{\partial R_{\ell^{\prime}}}%
+\Delta_{\ell^{\prime}}r_{\ell^{\prime}}\frac{\partial}{\partial r_{\ell}%
}+i\hbar\lambda_{22}\frac{\partial^{2}}{\partial r_{\ell}\partial
R_{\ell^{\prime}}}-i\frac{\Gamma_{\ell}}{\hbar}r_{\ell}R_{\ell^{\prime}%
}\right)  \right]  \widetilde{\rho}\text{,}%
\end{align}
where the diffusion coefficients $D_{\ell}$ and the effective coupling
parameters $\Gamma_{\ell}$ and $\Delta_{\ell}$ are given by
\begin{subequations}
\label{54}%
\begin{align}
D  &  =2m\widetilde{\gamma}k_{B}T\text{,}\\
\Gamma_{\ell}  &  =\lambda_{11}+2m\widetilde{\gamma}\sum_{\ell^{\prime}\left(
\neq\ell\right)  }\lambda_{\ell^{\prime}\ell}\text{,}\\
\Delta_{\ell}  &  =\left(  \lambda_{21}\delta_{\ell2}+\lambda_{12}\delta
_{\ell1}\right)  +2m\widetilde{\gamma}\lambda_{22}\text{.}%
\end{align}

Following the steps outlined above for the case of distinct reservoirs,
leading to the system of coupled ordinary differential equations (\ref{58}),
we now obtain the analogous system%

\end{subequations}
\begin{subequations}
\label{58}%
\begin{align}
\frac{\operatorname*{d}r_{\ell}}{\operatorname*{d}s}  &  =-\hbar\sum
_{\ell^{\prime}}\left[  \lambda_{22}\left(  1-\delta_{\ell\ell^{\prime}%
}\right)  +\frac{1}{m}\delta_{\ell\ell^{\prime}}\right]  K_{\ell^{\prime}%
}+2\widetilde{\gamma}r_{\ell}+\left(  \Delta_{1}r_{1}\delta_{\ell2}+\Delta
_{2}r_{2}\delta_{\ell1}\right)  \text{,}\\
\frac{\operatorname*{d}K_{\ell}}{\operatorname*{d}s}  &  =\frac{m\omega_{\ell
}^{2}r_{\ell}}{\hbar}+\frac{1}{\hbar}\left[  \left(  \Gamma_{2}r_{2}%
-\hbar\lambda_{12}K_{2}\right)  \delta_{\ell1}+\left(  \Gamma_{1}r_{1}%
-\hbar\lambda_{21}K_{1}\right)  \delta_{\ell2}\right]  \text{,}\\
\frac{\operatorname*{d}\widetilde{\rho}}{\operatorname*{d}s}  &  =-\frac
{1}{\hbar^{2}}D\left(  r_{1}+r_{2}\right)  ^{2}\widetilde{\rho}\text{,}\\
\frac{\operatorname*{d}t}{\operatorname*{d}s}  &  =1\text{.}%
\end{align}
As before, the first four equations is given in the matrix form as%

\end{subequations}
\begin{equation}
\frac{\operatorname*{d}}{\operatorname*{d}t}\left(
\begin{array}
[c]{c}%
r_{1}\left(  t\right) \\
K_{1}\left(  t\right) \\
r_{2}\left(  t\right) \\
K_{2}\left(  t\right)
\end{array}
\right)  =\left(
\begin{array}
[c]{cccc}%
2\widetilde{\gamma} & -\frac{\hbar}{m} & \Delta_{2} & -\hbar\lambda_{22}\\
\frac{1}{\hbar}m\omega_{1}^{2} & 0 & \frac{1}{\hbar}\Gamma_{2} & -\lambda
_{12}\\
\Delta_{1} & -\hbar\lambda_{22} & 2\widetilde{\gamma} & -\frac{\hbar}{m}\\
\frac{1}{\hbar}\Gamma_{1} & -\lambda_{21} & \frac{1}{\hbar}m\omega_{2}^{2} & 0
\end{array}
\right)  \left(
\begin{array}
[c]{c}%
r_{1}\left(  t\right) \\
K_{1}\left(  t\right) \\
r_{2}\left(  t\right) \\
K_{2}\left(  t\right)
\end{array}
\right)  \text{.}%
\end{equation}

From here on, all the derivations performed earlier for the case of distinct
reservoirs can be followed in exactly the same way, leading to the same
structure as the solution given in Eq. (\ref{75}). The difference is that the
elements of the above square matrix lead to values of $\eta_{mn}$ that differ
from those in the case of distinct reservoirs. Moreover, the function
$Z\left(  \left\{  K_{\ell}\right\}  ,\left\{  r_{\ell}\right\}  ,t\right)  $
becomes
\begin{equation}
Z\left(  \left\{  K_{\ell}\right\}  ,\left\{  r_{\ell}\right\}  ,t\right)
=D\sum_{m,n}c_{m}\left(  t\right)  c_{n}\left(  t\right)  \frac{\left(
\eta_{1m}+\eta_{3m}\right)  \left(  \eta_{1n}+\eta_{3n}\right)  }{\Lambda
_{m}+\Lambda_{n}}\left(  1-\operatorname{e}^{-\left(  \Lambda_{m}+\Lambda
_{n}\right)  t}\right)  \text{,}%
\end{equation}
instead of the expression given in Eq. (\ref{66}). Therefore, in spite of the
solutions to the master equation having the same form, the evolution of the
reduced density matrices of the two dissipative oscillators must be quite
different for the cases of one common reservoir and two separate ones. We
finally point out that the expression for the decay or decoherence
$\mathcal{D}(t)$ of the off-diagonal peaks of the initial density operator
also has the same structure as that in Eq. (\ref{83}), the only difference
being the time-dependent function $\Gamma(t)$, which differs in the two cases
of a common and two separate reservoirs, again due to the values of $\eta
_{mn}$.

\section{Decoherence in the double Caldeira-Leggett model}

In this section we analyze the decoherence of three particular entangled
states prepared in both oscillators of the network at $t=0$. These
entanglements, derived from Eq. (\ref{68}) under the assumption that they all
have the same initial mean energy $\left\langle E\right\rangle =$
$\left\langle H_{S_{1}}+H_{S_{2}}\right\rangle $ and distance $d$ between the
two peaks in the positional $x_{1}x_{2}$-space, are given by%

\begin{subequations}
\label{Psi}%
\begin{align}
\Psi^{(1)}\left(  \left\{  x_{\ell}\right\}  ,0\right)   &  =\mathcal{N}%
^{(1)}\left\{  \exp\left[  -\left(  x_{1}+q_{1}^{(1)}\right)  ^{2}/\sigma
_{1}^{2}\right]  \exp\left[  -\left(  x_{2}-q_{2}^{(1)}\right)  ^{2}%
/\sigma_{1}^{2}\right]  \right. \nonumber\\
&  +\left.  \exp\left[  -\left(  x_{1}+q_{2}^{(1)}\right)  ^{2}/\sigma_{1}%
^{2}\right]  \exp\left[  -\left(  x_{2}-q_{1}^{(1)}\right)  ^{2}/\sigma
_{1}^{2}\right]  \right\}  \text{,}\label{Psia}\\
\Psi^{(2)}\left(  \left\{  x_{\ell}\right\}  ,0\right)   &  =\mathcal{N}%
^{(2)}\left\{  \exp\left[  -\left(  x_{1}-q_{1}^{(2)}\right)  ^{2}/\sigma
_{2}^{2}\right]  \exp\left[  -\left(  x_{2}-q_{2}^{(2)}\right)  ^{2}%
/\sigma_{2}^{2}\right]  \right. \nonumber\\
&  +\left.  \exp\left[  -\left(  x_{1}-q_{2}^{(2)}\right)  ^{2}/\sigma_{2}%
^{2}\right]  \exp\left[  -\left(  x_{2}-q_{1}^{(2)}\right)  ^{2}/\sigma
_{2}^{2}\right]  \right\}  \text{,}\label{Psib}\\
\Psi^{(3)}\left(  \left\{  x_{\ell}\right\}  ,0\right)   &  =\mathcal{N}%
^{(3)}\left\{  \exp\left[  -\left(  x_{1}+q_{1}^{(3)}\right)  ^{2}/\sigma
_{3}^{2}\right]  \exp\left[  -\left(  x_{2}+q_{2}^{(3)}\right)  ^{2}%
/\sigma_{3}^{2}\right]  \right. \nonumber\\
&  +\left.  \exp\left[  -\left(  x_{1}-q_{1}^{(3)}\right)  ^{2}/\sigma_{3}%
^{2}\right]  \exp\left[  -\left(  x_{2}-q_{2}^{(3)}\right)  ^{2}/\sigma
_{3}^{2}\right]  \right\}  \text{.} \label{Psic}%
\end{align}
The choice of equal mean energies $\left\langle E\right\rangle $ and distances
$d$ follows from the fact that the decoherence process depends on the energy,
the distances between the components of the superposition, and the damping
rate defining the system-reservoir coupling. It is well-known that the
decoherence time varies inversely with the energy, the distance $d$, and the
damping rate. The above states differ from each other only by the position of
their peaks in the $x_{1}x_{2}$-space. While both peaks of $\Psi^{(1)}\left(
\left\{  x_{\ell}\right\}  ,0\right)  $ ($\Psi^{(2)}\left(  \left\{  x_{\ell
}\right\}  ,0\right)  $) are positioned in the fourth (first) quadrant of the
$x_{1}x_{2}$-space, those of the state $\Psi^{(3)}\left(  \left\{  x_{\ell
}\right\}  ,0\right)  $ are positioned in the first and third quadrant.
Moreover, whereas $\Psi^{(2)}\left(  \left\{  x_{\ell}\right\}  ,0\right)  $
is obtained from $\Psi^{(1)}\left(  \left\{  x_{\ell}\right\}  ,0\right)  $ by
a rotation in $x_{1}x_{2}$-space, $\Psi^{(3)}\left(  \left\{  x_{\ell
}\right\}  ,0\right)  $ requires, apart from the rotation, a displacement
operation over $\Psi^{(1)}\left(  \left\{  x_{\ell}\right\}  ,0\right)  $.

We analyze the decoherence time of the above entanglements for the two cases
of a common and separate reservoirs. We observe, for comparison, that the
analysis of decoherence in Refs. \cite{Mickel-AP2,Mickel-AP1,Mickel}, centered
on absolute zero reservoirs, shows that the decoherence rate for the case of a
common reservoir is attenuated, compared to the case of distinct ones.
However, the present analysis is based on the opposite scenario of the
high-temperature limit, so that we do not expect to obtain similar results to
those in Refs. \cite{Mickel-AP2,Mickel-AP1,Mickel}.

As the decoherence rate $\mathcal{D}(t)$ is given by Eq. (\ref{83}), in Figs.
1, 2, and 3, we plot $\mathcal{D}(t)$ against the scaled time $\gamma_{1}t$
for the states $\Psi^{(1)}$, $\Psi^{(2)}$, and $\Psi^{(3)}$, respectively. In
parts (a) and (b) of Figs. 1 and 2 we plot the decoherence rate $\mathcal{D}%
(t)$ for the cases of distinct reservoirs and a common one, respectively.
Adopting unit constants $\hbar,k_{B}=1$, masses ($m_{1}=m_{2}=1$), frequency
$\omega_{1}=1$ and widths $\sigma_{1}=\sigma_{2}=1$, we have set the
magnitudes $T_{1}=T_{2}=10^{3}\hbar\omega_{1}/k_{B}$ and $q_{1}^{(1)}%
=q_{2}^{(2)}=2q_{2}^{(1)}$ $=2q_{1}^{(2)}=10\sigma_{1}$. We have also assumed
a regime of parameters where $\gamma_{1}=\gamma_{2}\ll\left\{  \lambda
_{\ell\ell^{\prime}}\right\}  \,<\omega_{1}=\omega_{2}/2$, with $\gamma
_{1}/\omega_{1}=10^{-3}$. With these values, the mean energy $\left\langle
E\right\rangle $ and distance $d$ associated with states (\ref{Psi}) becomes
$\left\langle E\right\rangle \approx158\hbar\omega_{1}$ and $d=5\sqrt{2}%
\sigma_{1}$. To reach the same $\left\langle E\right\rangle $ and $d$ for all
three states we have assumed that $\sigma_{1}=12\sigma_{3}$, apart from the
relation $q_{1}^{(3)}=\sqrt{3/2}q_{2}^{(3)}=\sqrt{15/2}\sigma_{1}$.

For comparison, the thick solid line in all three figures represents the
decoherence time of the Schr\"{o}dinger-cat-like state
\end{subequations}
\begin{equation}
\Psi\left(  x,0\right)  =\mathcal{N}\left\{  \exp\left[  -\left(  x+q\right)
^{2}/\sigma^{2}\right]  +\exp\left[  -\left(  x-q\right)  ^{2}/\sigma
^{2}\right]  \right\}  \text{,} \label{SCLS}%
\end{equation}
prepared in\ one of the oscillators, decoupled from the other. This
Schr\"{o}dinger-cat-like state also leads to the same values established above
for the mean energy $\left\langle E\right\rangle $ and distance $d$ between
the two peaks in the $x$-space. To achieve this, we set the relations $q=$
$5\sigma_{1}/\sqrt{2}$and $\sigma=6\times10^{-2}\sigma_{1}$. Therefore, the
thick solid line describes the decoherence process of a superposition state in
the CL problem.

In all three figures, the solid (dashed) and dashed-dotted (dotted) lines
describe the decoherence processes when considering distinct reservoirs (a
common one) and the coupling between the oscillators given by $\lambda
_{11}q_{1}q_{2}+\lambda_{22}p_{1}p_{2}$ and $\lambda_{12}q_{1}p_{2}%
+\lambda_{21}q_{2}p_{1}$, respectively, with $\lambda_{\ell\ell^{\prime}}%
=0.1$. Both couplings $\lambda_{11}q_{1}q_{2}+\lambda_{22}p_{1}p_{2}$ and
$\lambda_{12}q_{1}p_{2}+\lambda_{21}q_{2}p_{1}$, when described in terms of
the usual annihilation (creation) operators $a_{1}$,$a_{2}$ ($a_{1}^{\dagger}%
$,$a_{2}^{\dagger}$), correspond to the rotating terms $a_{1}^{\dagger}%
a_{2}+a_{1}a_{2}^{\dagger}$ and the counter-rotating terms $i\left(
a_{1}a_{2}-a_{1}^{\dagger}a_{2}^{\dagger}\right)  $, respectively. We observe
that the decay rates of the curves in Figs. 1 and 2 (a and b) are around that
associated with the Schr\"{o}dinger-cat-like state for both cases of a common
and distinct reservoirs. Therefore, the case of a common reservoir does not
exhibit advantages over that of distinct reservoirs, as demonstrated
previously for absolute zero reservoirs \cite{Mickel-AP2,Mickel-AP1,Mickel}.

Considering now Fig. 3, we observe that all curves decay faster than those in
Figs. 1 and 2. Moreover, the decay rates of the curves associated with the
coupling $\lambda_{11}q_{1}q_{2}+\lambda_{22}p_{1}p_{2}$ are even faster than
those associated with $\lambda_{12}q_{1}p_{2}+\lambda_{21}q_{2}p_{1}$. This
behavior follows from Eq. (\ref{40}), which shows that the larger the coupling
strength $\lambda_{22}$, the larger the effective damping rate $\widetilde
{\gamma}_{\ell}$. The same explanation applies to Fig. 4, where the same
curves as in Fig. 1 are plotted for larger strengths $\lambda_{\ell
\ell^{\prime}}=0.5$; we observe that the decay rates of the curves derived
from the coupling $\lambda_{11}q_{1}q_{2}+\lambda_{22}p_{1}p_{2}$ are
significantly faster than those for $\lambda_{12}q_{1}p_{2}+\lambda_{21}%
q_{2}p_{1}$, as in Fig. 3. In Figs. 1 and 2, the effect coming from the
relation between $\lambda_{22}$ and $\widetilde{\gamma}_{\ell}$ is blurred,
making the decay rates of the curves associated with the coupling
$\lambda_{11}q_{1}q_{2}+\lambda_{22}p_{1}p_{2}$ similar to those for
$\lambda_{12}q_{1}p_{2}+\lambda_{21}q_{2}p_{1}$.

Differently from the works in Refs. \cite{Mickel-AP2,Mickel-AP1,Mickel}, where
the decay rates were chiefly determined by the assumption of one common or two
separate reservoirs, here the difference between the decay rates comes from
the different coupling mechanisms between the oscillators, $\lambda_{11}%
q_{1}q_{2}+\lambda_{22}p_{1}p_{2}$ or $\lambda_{12}q_{1}p_{2}+\lambda
_{21}q_{2}p_{1}$. The reason for this is that Refs.
\cite{Mickel-AP2,Mickel-AP1,Mickel} apply to absolute zero reservoirs, where
the coupling between the oscillators induced by a common reservoir tends to
decrease the decoherence rates. The same effect of coherence control is also
achieved in Refs. \cite{Mickel-AP2,Mickel-AP1,Mickel} by assuming that the
network oscillators are strongly coupled to each other. Especially in Ref.
\cite{Mickel-AP2}, it is demonstrated that the R-DFSs emerge from situations
where the whole network interacts with a common reservoir or when each
resonator, strongly coupled to each other, interacts with its own reservoir.
The present work, however, applies to the high-temperature regime where, as
demonstrated in Figs. 1, 2 and 3, the interaction induced by a common
reservoir is completely blurred by the high-temperature effects. From this
fact we may expect the high-temperature regime to prevent the emergence of
R-DFSs, at least in a network with a small number of oscillators, as in the
case at hand.

\section{Concluding Remarks}

In this study we analyze the double CL model, i.e., the path integral approach
to two interacting dissipative harmonic oscillators. We derive and solve the
master equations associated with two different situations: when each
oscillator is coupled to its own reservoir, and when both oscillators are
coupled to a common reservoir. In both cases, the derived master equations
consist of two CL models --- describing the dissipative mechanism of each
oscillator independently --- as well as the dynamics arising from the
interaction between the two oscillators. However, in the case of a common
reservoir, we identify a reservoir-induced coupling between the oscillators,
even when the original interaction between them is schwitched off. Such a
reservoir-induced coupling, recently pointed out in Ref. \cite{Amir},
encompasses both dissipative and diffusive terms which couple together the
variables of both oscillators. These terms thus account for the energy loss of
the oscillators through each other, apart from a joint diffusive process.

The occurrence of such a reservoir-induced coupling between the oscillators
was also pointed out in Refs. \cite{Mickel-AP2,Mickel-AP1,Mickel,MickelRG},
where networks of dissipative quantum harmonic oscillators are treated through
perturbative master equations. In Refs.
\cite{Mickel-AP2,Mickel-AP1,Mickel,MickelRG}, the occurrence of indirect-decay
channels --- by which the network oscillators lose excitation through all the
other oscillators --- is demonstrated in two different situations: when all
the (non-interacting or interacting) oscillators are coupled to a common
reservoir and even when strongly interacting oscillators are coupled to their
own reservoirs.

Regarding the master equation solutions reached in this paper, we stress that
their form enables a full comprehension of the evolution of initial states of
the network, enlarging the perspective of the coherence and decoherence
analysis offered by the CL model \cite{PRA}. Through these solutions we
compute a general expression for the decay rate of the off-diagonal peaks of
the density matrix of initial superposition states, which applies to both
cases of a common and distinct reservoirs. Such an expression offers complete
information about the decoherence of the initial state of the coupled
dissipative oscillators. In this regard, we have analyzed, as an application,
the decoherence process of particular entanglements in the positional space of
both oscillators. The results demonstrated that the coupling induced by the
common reservoir does not lead to the collective damping effect mentioned
above. The high-temperature regime of validity for our calculations completely
blurred such reservoir-induced coupling which at absolute zero works, in
general, to delay the decoherence process or even to produce the R-DFSs.
However, we find that different interactions between the dissipative
oscillators, described by rotating or counter-rotating terms, result in
different decay rates of the interference terms of the density matrix. The
decay rates associated with the counter-rotating terms of the interaction
between the oscillators are significantly faster than those coming from the
rotating terms. The reason for this is that the effective damping constants
increase with increasing coupling strength related to the counter-rotating terms.

We note that a recent paper addressed the question of the derivation of a
master equation for two coupled harmonic oscillators through the
influence-functional method of Feynman and Vernon \cite{Hu}. However, the
authors assume a coupling between the oscillators different from that given by
Eq. (\ref{4}), apart from considering only the case of a common reservoir. We
finally stress that the present double CL model can be used for the analysis
of dissipative bipolarons, besides other problems in several areas of physics
where the CL model has been successfully employed over the last few decades. A
more detailed analysis of the development in Ref. \cite{Amir} is one immediate
application of the present study. The effect of temperature on decoherence and
the emergence of R-DFSs in the low-temperature regime is also a point worth
looking into. Furthermore, the tunneling process of coupled dissipative
systems is a phenomenon which may be accounted for by the present work.
Together with the achievements in Refs.
\cite{Mickel-AP2,Mickel-AP1,Mickel,MickelRG}, we believe that the present work
furnishes a great deal of material for discussion of the physics of coupled
dissipative systems.

\section{Appendix A - Diagonalization of the Hamiltonian $H_{S_{1}+S_{2}}$}

The Hamiltonian modeling the two harmonic oscillators and their mutual
interaction, $H_{S_{1}+S_{2}}$, as given by Eq. (\ref{10.5}) with ${V_{\ell}%
}(q_{\ell})=m_{\ell}\omega_{\ell}^{2}q_{\ell}^{2}/2$, can be rewritten in
terms of the quantum mechanical creation $a_{\ell}^{\dag}$ and annihilation
$a_{\ell}$ operators, as%

\begin{equation}
H_{S_{1}+S_{2}}=\hbar\left[  \sum_{\ell}\omega_{\ell}\left(  a_{\ell}^{\dag
}a_{\ell}+\frac{1}{2}\right)  +\left(  g_{1}a_{1}a_{2}+g_{2}a_{1}%
a_{2}^{\dagger}+H.c\right)  \right]  \text{,} \label{A1}%
\end{equation}
with the coupling strength%

\begin{equation}
g_{\ell}=\frac{\lambda_{11}-i\lambda_{21}m_{1}\omega_{1}}{\sqrt{m_{1}%
\omega_{1}m_{2}\omega_{2}}}+\left(  -1\right)  ^{\ell}\sqrt{m_{1}\omega
_{1}m_{2}\omega_{2}}\left(  \lambda_{22}+i\frac{\lambda_{12}}{m_{1}\omega_{1}%
}\right)  \text{.} \label{A2}%
\end{equation}
The diagonalization of the form (\ref{A1}), easily performed than that in Eq.
(\ref{10.5}), leads to $H_{S_{1}+S_{2}}=\hbar\sum_{\ell}\Omega_{\ell}\left(
A_{\ell}^{\dag}A_{\ell}+\frac{1}{2}\right)  $, with the normal-mode frequencies,%

\begin{equation}
\Omega_{\ell}^{2}=\frac{\omega_{1}^{2}+\omega_{2}^{2}}{2}-\left\vert
g_{1}\right\vert ^{2}+\left\vert g_{2}\right\vert ^{2}-(-1)^{\ell}%
\sqrt{\left(  \frac{\omega_{1}^{2}-\omega_{2}^{2}}{2}\right)  ^{2}-\left\vert
g_{1}\right\vert ^{2}\left(  \omega_{1}-\omega_{2}\right)  ^{2}+\left\vert
g_{2}\right\vert ^{2}\left(  \omega_{1}+\omega_{2}\right)  ^{2}}\text{,}
\label{A3}%
\end{equation}
and the normal-mode operators

\begin{equation}
A_{\ell}=\mathcal{N}_{\ell}\sum_{\ell^{\prime}}\left[  \Delta_{1\ell^{\prime}%
}(\Omega_{\ell})a_{\ell^{\prime}}^{\dagger}+\Delta_{2\ell^{\prime}}%
(\Omega_{\ell})a_{\ell^{\prime}}\right]  \text{,} \label{A4}%
\end{equation}
where
\begin{subequations}
\label{A5}%
\begin{align}
\mathcal{N}_{\ell^{\prime\prime}}^{-2}  &  =\sum\limits_{\ell,\ell^{\prime}%
}(-1)^{\ell}\Delta_{\ell\ell^{\prime}}^{2}(\Omega_{\ell^{\prime\prime}%
})\text{,}\label{A5a}\\
\Delta_{\ell\ell^{\prime}}\left(  \Omega_{\ell^{\prime\prime}}\right)   &
=2\delta_{\ell^{\prime}1}\left\vert g_{2}g_{\ell-(-1)^{\ell}}\right\vert
\omega_{2}+\left[  \delta_{\ell^{\prime}2}\left\vert g_{\ell-(-1)^{\ell}%
}\right\vert -\delta_{\ell1}\delta_{\ell^{\prime}1}\left(  \Omega
_{\ell^{\prime\prime}}-\omega_{2}\right)  \right] \nonumber\\
&  \times\left[  \left\vert g_{1}\right\vert ^{2}-\left\vert g_{2}\right\vert
^{2}+\left(  \Omega_{\ell^{\prime\prime}}-\omega_{1}\right)  (\Omega
_{\ell^{\prime\prime}}+(-1)^{\ell+\ell^{\prime}}\omega_{2})\right]  \text{.}
\label{A5b}%
\end{align}
The normal-mode coordinates $Q_{\ell}$ and $P_{\ell}$, following from the
operators $A_{\ell}$ and $A_{\ell}^{\dagger}$, are given by
\end{subequations}
\begin{subequations}
\label{A6}%
\begin{align}
Q_{\ell}  &  =\sqrt{\frac{2\hslash}{\Omega_{\ell}}}\sum\limits_{\ell^{\prime}%
}\mathrm{\operatorname{Re}}(a_{\ell\ell^{\prime}}q_{\ell^{\prime}}+b_{\ell
\ell^{\prime}}p_{\ell^{\prime}})\text{,}\label{A6a}\\
P_{\ell}  &  =\sqrt{2\hslash\Omega_{\ell}}\sum\limits_{\ell^{\prime}%
}\mathrm{\operatorname{Im}}(a_{\ell\ell^{\prime}}q_{\ell^{\prime}}+b_{\ell
\ell^{\prime}}p_{\ell^{\prime}})\text{,} \label{A6b}%
\end{align}
with the coefficients
\end{subequations}
\begin{subequations}
\label{A7}%
\begin{align}
a_{\ell^{\prime}\ell}  &  =\mathcal{N}_{\ell}\sqrt{\frac{m_{\ell}\omega_{\ell
}}{2\hslash}}\sum\limits_{\ell^{\prime\prime}}\Delta_{\ell^{\prime\prime}\ell
}\left(  \Omega_{\ell^{\prime}}\right)  \operatorname{e}^{i\left(  -1\right)
^{\ell^{\prime\prime}}\phi_{\ell}}\text{,}\label{A7a}\\
b_{\ell^{\prime}\ell}  &  =i\mathcal{N}_{\ell}\sqrt{\frac{1}{2\hslash m_{\ell
}\omega_{\ell}}}\sum\limits_{\ell^{\prime\prime}}\left(  -1\right)
^{\ell^{\prime\prime}}\Delta_{\ell^{\prime\prime}\ell}\left(  \Omega
_{\ell^{\prime}}\right)  \operatorname{e}^{i\left(  -1\right)  ^{\ell
^{\prime\prime}}\phi_{\ell}}\text{,} \label{A7b}%
\end{align}
and phase factors $\phi_{\ell}=\left[  \theta_{1}-(-1)^{\ell}\theta
_{2}\right]  /2$, where the quantities%
\end{subequations}
\begin{equation}
\theta_{\ell}=\cos^{-1}\left\{  \frac{\xi_{11}+(-1)^{\ell}\xi_{22}}{\left[
\xi_{11}+(-1)^{\ell}\xi_{22}\right]  ^{2}\left[  \xi_{12}+\xi_{21}\right]
^{2}}\right\}  \text{,}%
\end{equation}
are defined by the dimensionless strengths
\begin{subequations}
\label{A9}%
\begin{align}
\xi_{\ell\ell}  &  =\frac{\lambda_{\ell\ell}\left[  \delta_{\ell1}+\left(
m_{1}m_{2}\omega_{1}\omega_{2}\right)  ^{1/2}\delta_{\ell2}\right]  }{2\left[
\left(  m_{1}m_{2}\omega_{1}\omega_{2}\right)  ^{1/2}\delta_{\ell1}%
+\delta_{\ell2}\right]  }\text{,}\label{A9a}\\
\xi_{\ell\ell^{\prime}}  &  =\frac{\lambda_{\ell\ell^{\prime}}}{2}\sqrt
{\frac{m_{\ell}\omega_{\ell}}{m_{\ell^{\prime}}\omega_{\ell^{\prime}}}%
}\text{.} \label{A9b}%
\end{align}

Through the coordinates $Q_{\ell}$ and $P_{\ell}$, we finally obtain the
diagonalized Hamiltonian%

\end{subequations}
\begin{equation}
\mathcal{H}_{S_{1}+S_{2}}=\frac{1}{2}\sum\limits_{\ell}\left(  P_{\ell}%
^{2}+\Omega_{\ell}^{2}Q_{\ell}^{2}\right)  \text{.} \label{A8}%
\end{equation}

\textbf{Acknowledgments}

We wish to express thanks for the support from FAPESP and CNPq, Brazilian agencies.

\textbf{Figure captions}

Fig. 1 Plot of the decay or decoherence function $\mathcal{D}(t)$ against the
scaled time $\gamma_{1}t$ for the states $\Psi^{(1)}$ in Eq. (\ref{Psia}),
considering the cases of (a) distinct reservoirs and (b) a common one. The
thick solid line describes the decoherence process of the
Schr\"{o}dinger-cat-like state $\Psi\left(  x,0\right)  $ in Eq. (\ref{SCLS}),
prepared in a single dissipative oscillator. The solid (dashed) and
dashed-dotted (dotted) lines describe the decoherence processes of $\Psi
^{(1)}$ when assuming distinct reservoirs (a common one) and the coupling
between the oscillators given by $\lambda_{11}q_{1}q_{2}+\lambda_{22}%
p_{1}p_{2}$ and $\lambda_{12}q_{1}p_{2}+\lambda_{21}q_{2}p_{1}$, respectively,
with $\lambda_{\ell\ell^{\prime}}=0.1$.

Fig. 2 Plot of the decay or decoherence function $\mathcal{D}(t)$ against the
scaled time $\gamma_{1}t$ for the states $\Psi^{(2)}$ in Eq. (\ref{Psib}),
considering the cases of (a) distinct reservoirs and (b) a common one. The
thick solid line describes the decoherence process of the
Schr\"{o}dinger-cat-like state $\Psi\left(  x,0\right)  $ in Eq. (\ref{SCLS}),
prepared in a single dissipative oscillator. The solid (dashed) and
dashed-dotted (dotted) lines describe the decoherence processes of $\Psi
^{(2)}$ when assuming distinct reservoirs (a common one) and the coupling
between the oscillators given by $\lambda_{11}q_{1}q_{2}+\lambda_{22}%
p_{1}p_{2}$ and $\lambda_{12}q_{1}p_{2}+\lambda_{21}q_{2}p_{1}$, respectively,
with $\lambda_{\ell\ell^{\prime}}=0.1$.

Fig. 3 Plot of the decay or decoherence function $\mathcal{D}(t)$ against the
scaled time $\gamma_{1}t$ for the states $\Psi^{(3)}$ in Eq. (\ref{Psic}),
considering the cases of distinct reservoirs and a common one. The thick solid
line describes the decoherence process of the Schr\"{o}dinger-cat-like state
$\Psi\left(  x,0\right)  $ in Eq. (\ref{SCLS}), prepared in a single
dissipative oscillator. The solid (dashed) and dashed-dotted (dotted) lines
describe the decoherence processes of $\Psi^{(3)}$ when assuming distinct
reservoirs (a common one) and the coupling between the oscillators given by
$\lambda_{11}q_{1}q_{2}+\lambda_{22}p_{1}p_{2}$ and $\lambda_{12}q_{1}%
p_{2}+\lambda_{21}q_{2}p_{1}$, respectively, with $\lambda_{\ell\ell^{\prime}%
}=0.1$.

Fig. 4 Plot of the decay or decoherence function $\mathcal{D}(t)$ against the
scaled time $\gamma_{1}t$ for the states $\Psi^{(1)}$ in Eq. (\ref{Psia}),
considering the cases of (a) distinct and (b) a common reservoir. The thick
solid line describes the decoherence process of the Schr\"{o}dinger-cat-like
state $\Psi\left(  x,0\right)  $ in Eq. (\ref{SCLS}), prepared in a single
dissipative oscillator. The solid (dashed) and dashed-dotted (dotted) lines
describe the decoherence processes of $\Psi^{(1)}$ when assuming distinct
reservoirs (a common one) and the coupling between the oscillators given by
$\lambda_{11}q_{1}q_{2}+\lambda_{22}p_{1}p_{2}$ and $\lambda_{12}q_{1}%
p_{2}+\lambda_{21}q_{2}p_{1}$, respectively, with $\lambda_{\ell\ell^{\prime}%
}=0.5$.

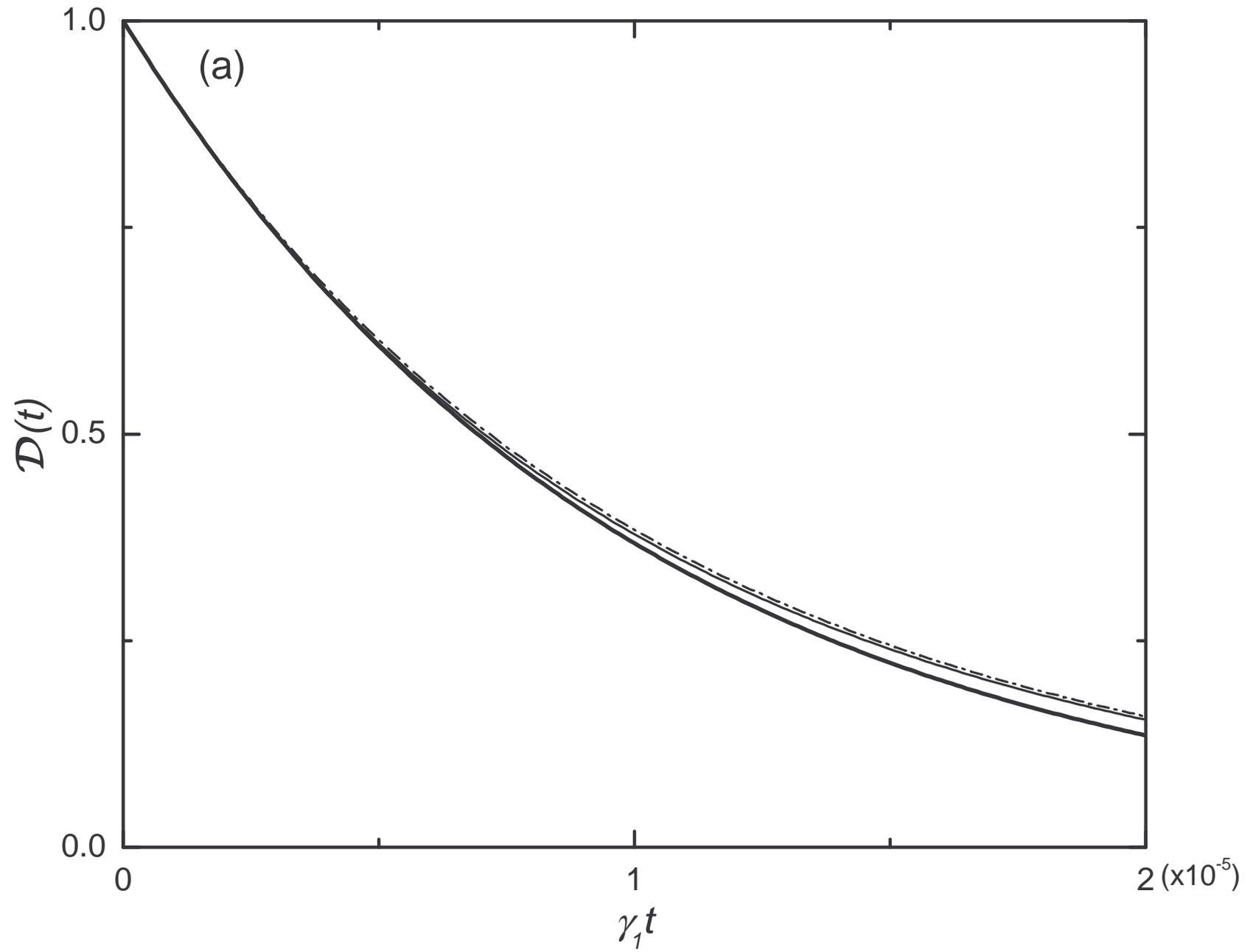
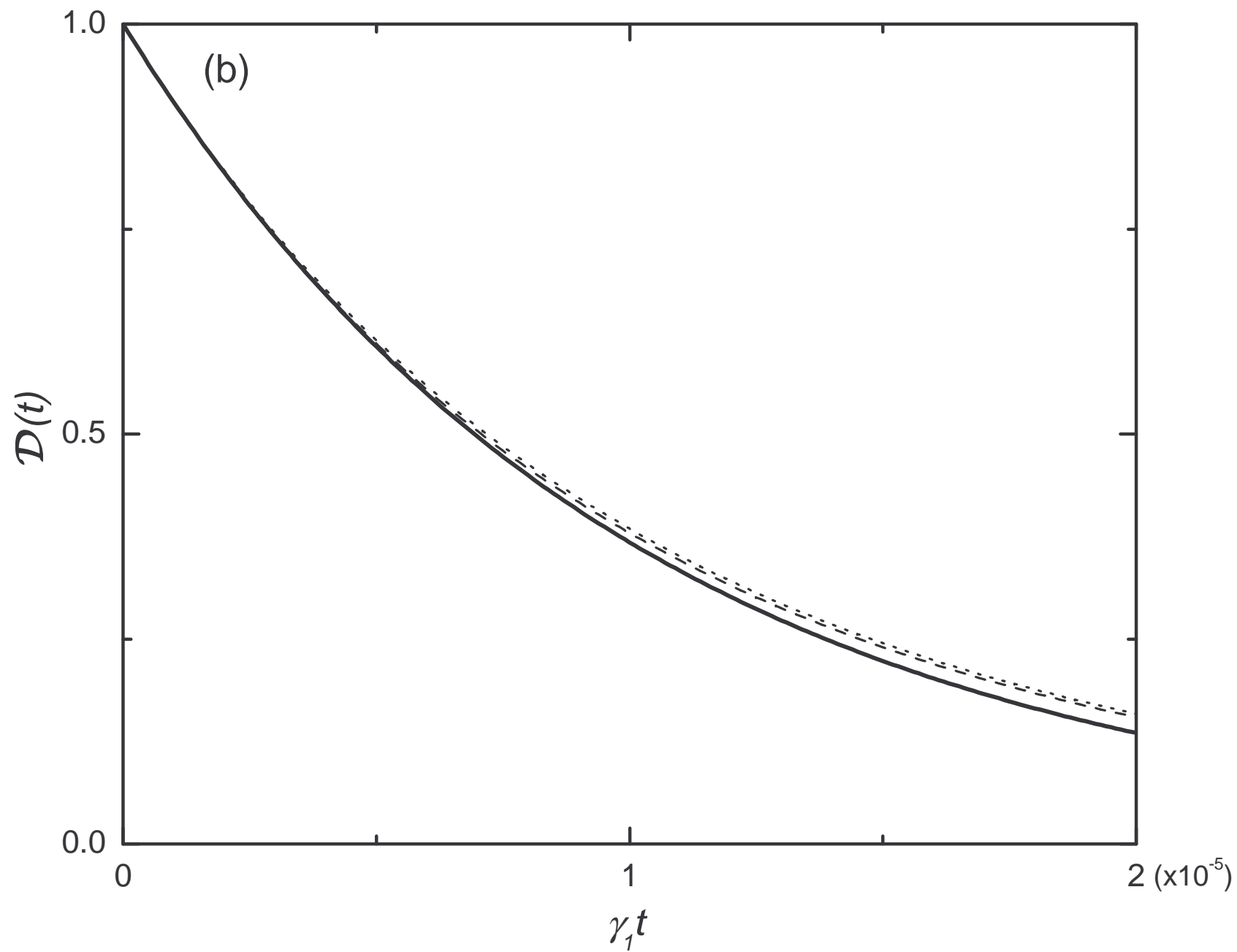
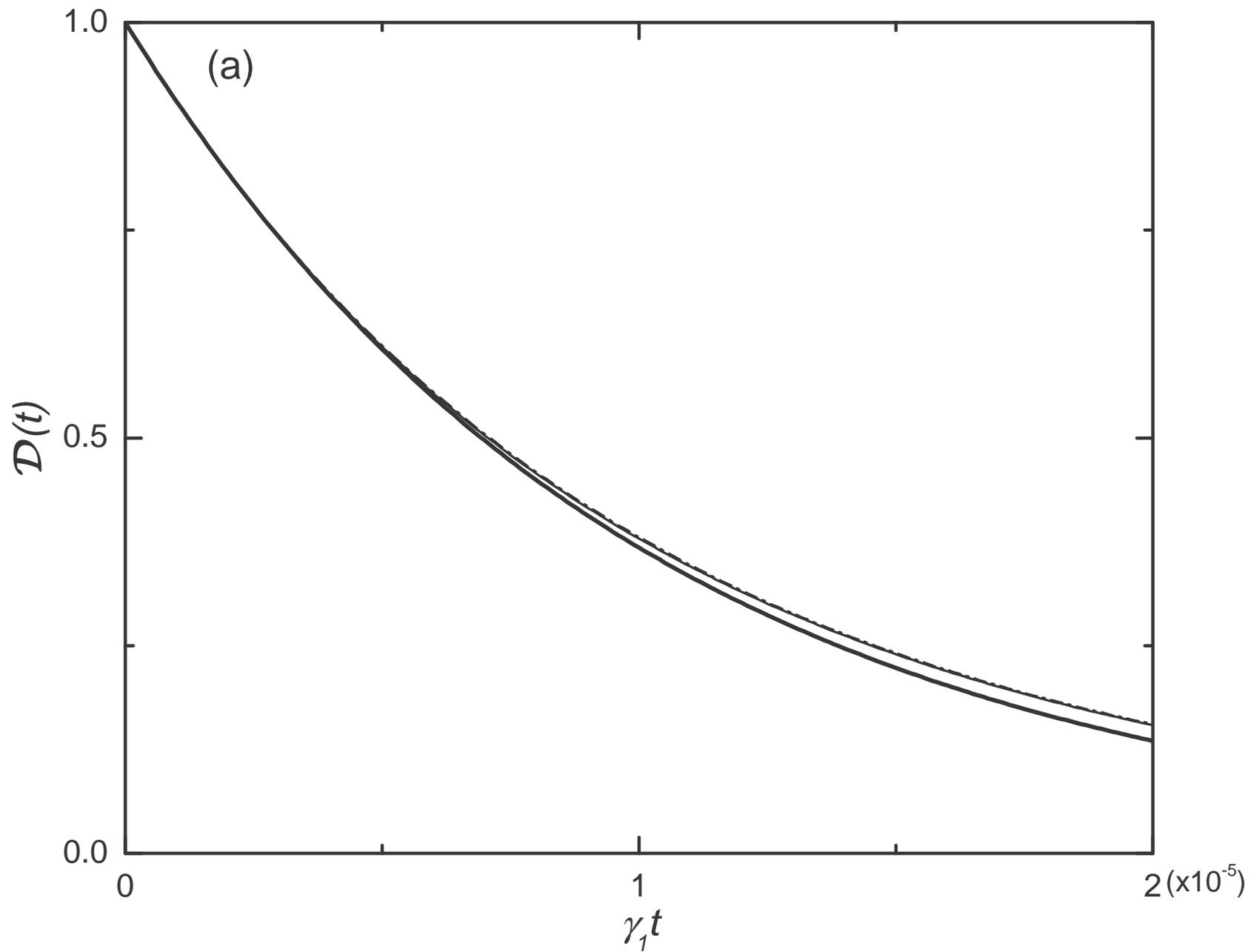
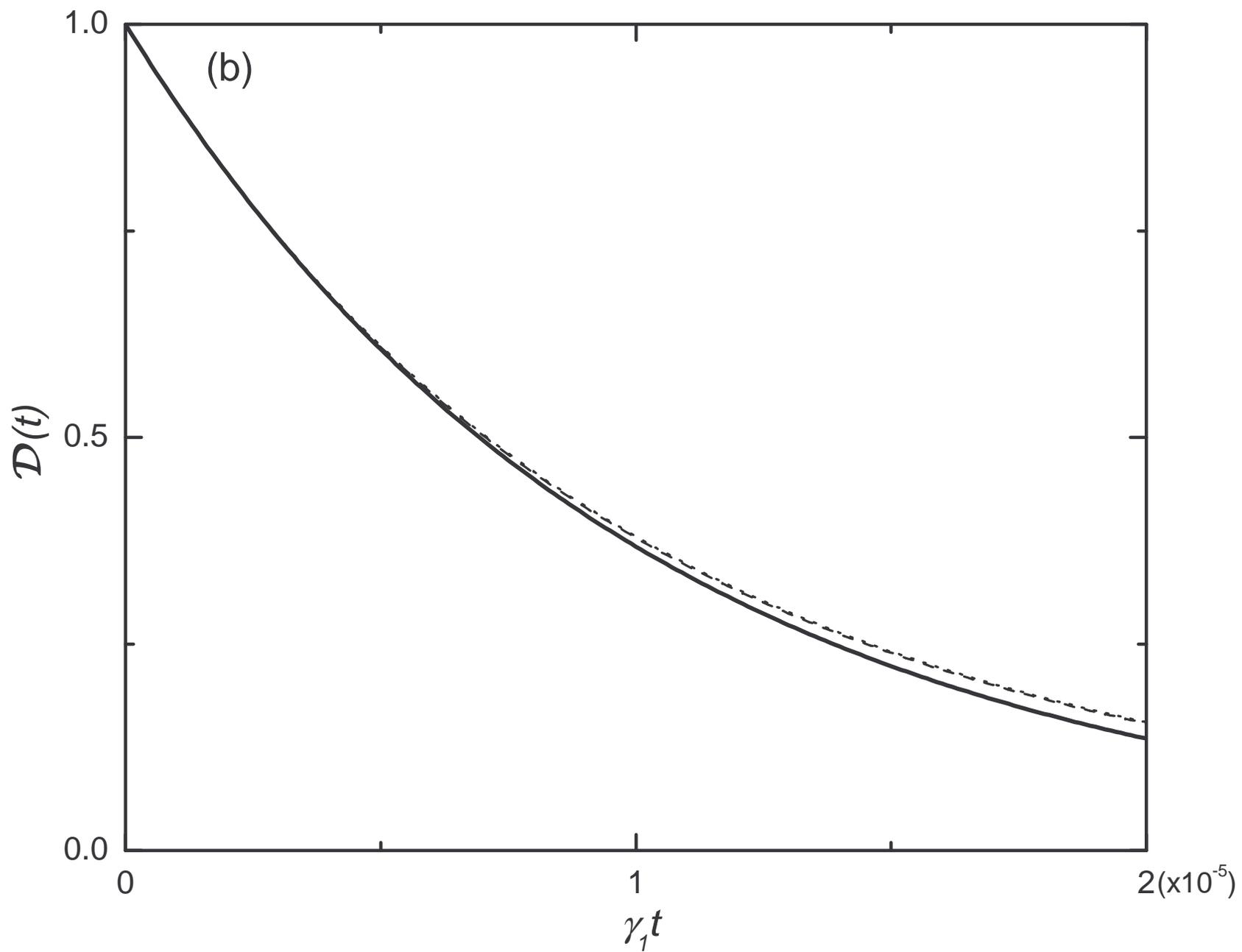
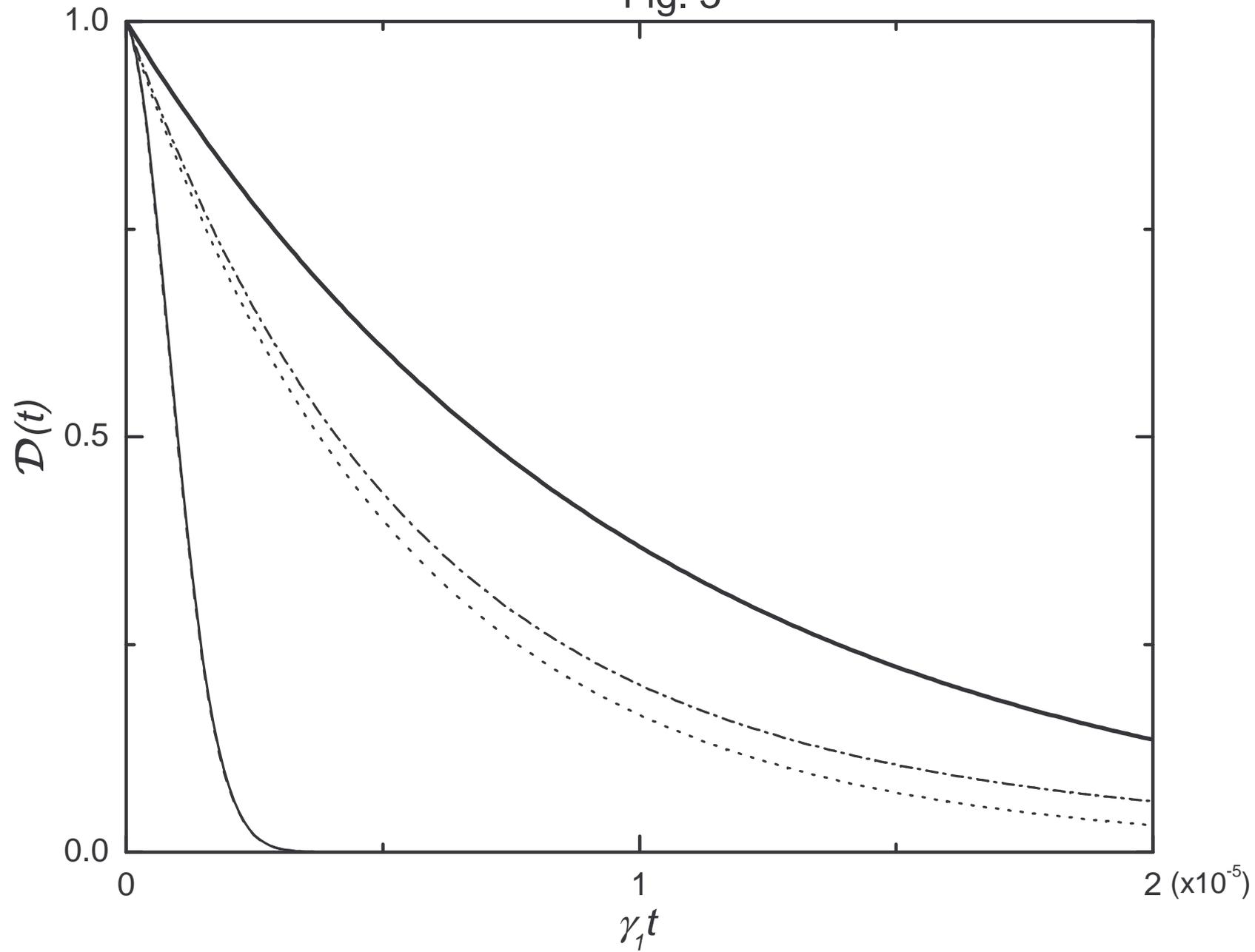
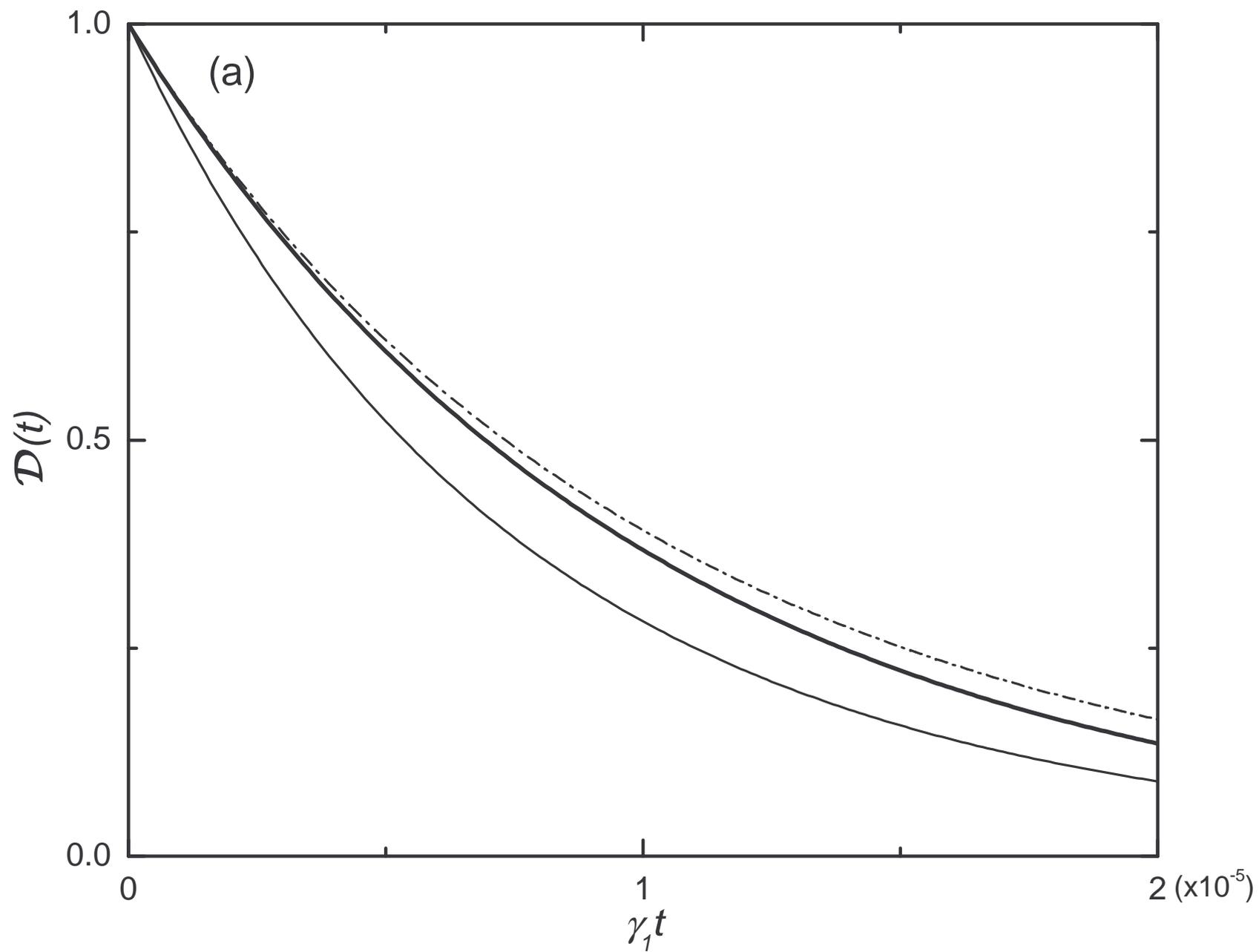
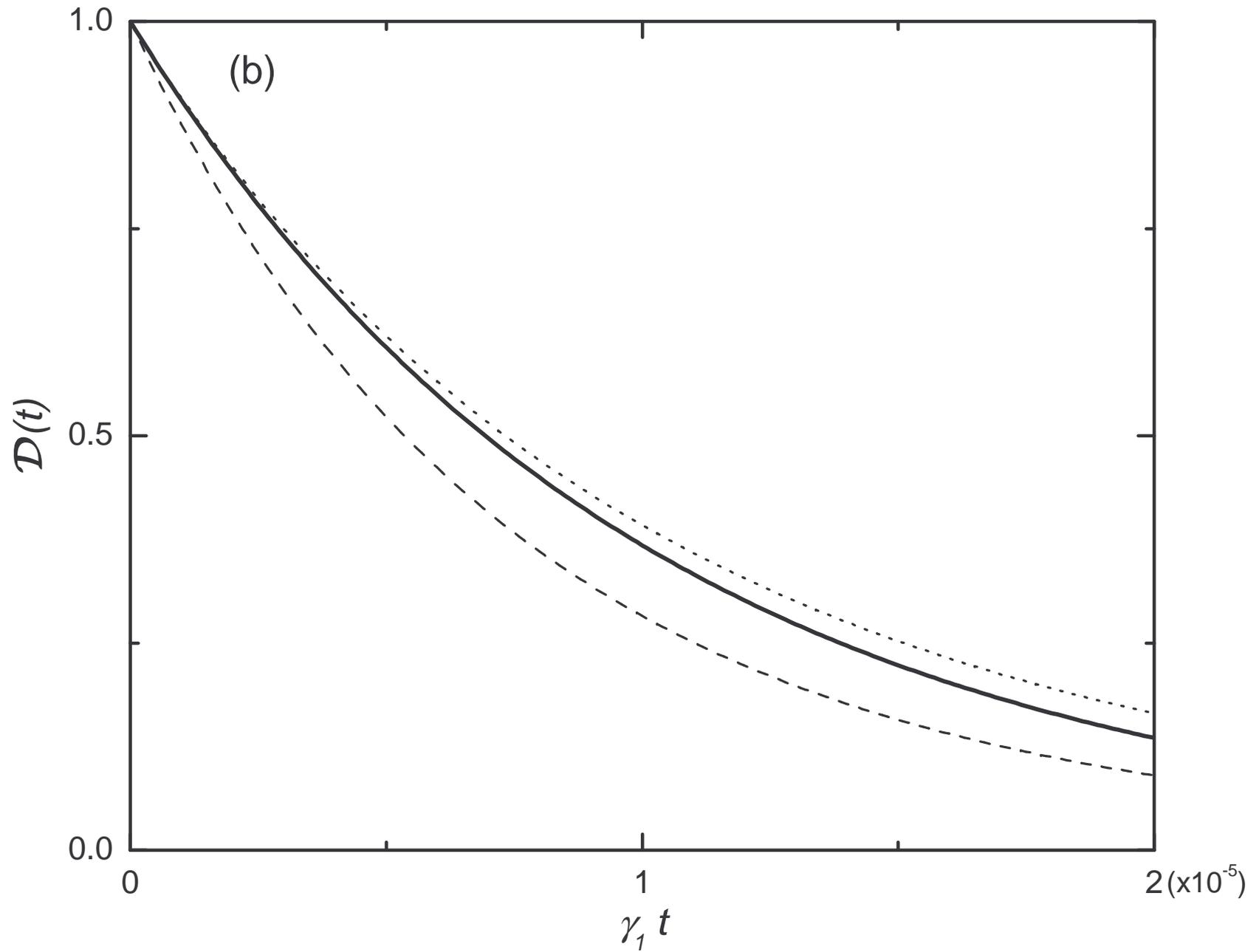
\end{document}